\documentclass[final,1p,times]{elsarticle}
\usepackage{here}

\usepackage{amsmath,amssymb,bm}
\usepackage{comment}

\begin{document}

\begin{flushright}
KEK-TH-2309 and J-PARK-TH-0239
\end{flushright}

\begin{frontmatter}




\title{Revisiting the compatibility problem between the gauge principle and 
the observability of the canonical orbital angular momentum in 
the Landau problem}


\author[label1,label2]{Masashi~Wakamatsu\corref{cor1}}
\ead{wakamatu@post.kek.jp}
\author[label3]{Yoshio~Kitadono}
\author[label4,label2]{Liping~Zou}
\author[label5]{Pengming~Zhang}

\cortext[cor1]{corresponding author}

\address[label1]{KEK Theory Center, Institute of Particle and Nuclear Studies,\\
High Energy Accelerator Research Organization (KEK),
1-1, Oho, Tsukuba, Ibaraki 305-0801, Japan}
\address[label2]{Institute of Modern Physics, Chinese Academy of Sciences,
Lanzhou, People's Republic of China, 730000}
\address[label3]{Liberal Education Center, National Chin-Yi University of Technology,
No.57, Sec.2, \\
Zhongshan Rd., Taiping Dist., Taichung 41170, Taiwan, R.O.C.}
\address[label4]{Sino-French Institute of Nuclear Engineering and Technology, 
Sun Yat-Sen University, Zhuhai 519082, \\
People's Republic of China} 
\address[label5]{School of Physics and Astronomy, Sun Yat-sen University,
Zhuhai, 519082, People's Republic of China}

\begin{abstract}
As is widely-known, the eigen-functions of the Landau problem in the
symmetric gauge are specified by two quantum numbers. 
The first is the familiar Landau quantum number $n$, whereas the 
second is the magnetic quantum number $m$, which is the eigen-value 
of the canonical orbital angular momentum (OAM) operator of the electron. 
The eigen-energies of the system depend only on the first quantum 
number $n$, and the second quantum number $m$ does not correspond 
to any direct observables. This seems natural since the canonical OAM 
is generally believed to be a {\it gauge-variant} quantity, and observation 
of a gauge-variant quantity would contradict a fundamental principle of 
physics called the {\it gauge principle}.
In recent researches, however, Bliokh et al. analyzed the motion of 
helical electron beam along the direction of a uniform magnetic field,
which was mostly neglected in past analyses of the
Landau states. Their analyses revealed highly non-trivial $m$-dependent
rotational dynamics of the Landau electron, but the problem is that
their papers give an impression that the quantum number $m$ in the
Landau eigen-states corresponds to a genuine observable. 
This compatibility problem between the gauge principle and the observability 
of the quantum number $m$ in the Landau eigen-states was attacked
in our previous letter paper. In the present paper, we try to give 
more convincing answer to this delicate problem of physics, especially by
paying attention not only to the {\it particle-like} aspect but also to the 
{\it wave-like} aspect of the Landau electron.
\end{abstract}


\end{frontmatter}



\section{Introduction}
\label{Section1}

Under the presence of magnetic field background, the
canonical momentum as well as the canonical orbital angular
momentum (OAM) of a charged particle is believed to be
gauge-variant quantities different from the mechanical
(or kinetic) momentum and the mechanical OAM.
Because of their gauge-variant nature, the canonical
quantities are generally believed not to correspond to
observables. In fact, an observation of the gauge-variant
quantity would contradict the so-called gauge principle as
one of the important principles of physics.
To clearly understand the difference between the canonical
OAM and the mechanical OAM is also a key issue for judging
which of the two types of decomposition, i.e. the canonical
type and of the mechanical type is favored from the physical
perspective in the so-called nucleon spin decomposition 
problem \cite{JM1990}\nocite{Ji1997}\nocite{BJ1998}
\nocite{Waka2010}\nocite{Waka2011}\nocite{Burkardt2013}
\nocite{JZZ2015}\nocite{Waka2015}-\cite{Waka2016}.
(For review of the nucleon spin decomposition problem,
see \cite{Review_LL2014},\cite{Review_Waka2014}.)
We have already tried to clarify the meanings and the differences of
these two different OAMs thorough the studies of the familiar 
Landau problem \cite{WKZ2018} as well as of the Aharonov-Bohm 
effect \cite{WKZZ2018}. In the present paper,
we extend these analyses to the physics of helical electron beam in a
uniform magnetic field, with a particular intention of resolving the
compatibility problem between the gauge principle and the observability
of the canonical OAM. (A part of this investigation was 
already reported in a letter paper \cite{WKZZ2020}. One of the 
interesting topics, which was discussed in this previous paper
but is left out in the present paper, is an intuitive or physical 
explanation of the $m$-dependent splitting of the helical electron 
beam based on the novel concept of {\it quantum guiding center} 
in the Landau problem \cite{KWZZ2020},\cite{Enk2020}. )

The existence of propagating wave carrying intrinsic orbital angular
momentum (OAM) has been an object of intensive study for many years
and it is firmly established by now not only for photon 
beams but also for electron beams \cite{ABSW1992}\nocite{APB1999}
\nocite{TT2011}\nocite{BBSN2007}-\cite{Breview2017}. 
These helical beams are
characterized by an integer $m$ sometimes called the topological index of
the twisted beam.
This integer is nothing but the eigen-value of the canonical
OAM operator, or more precisely its component along the propagating
direction of the photon or electron beam.
Note that the canonical OAM is standardly believed to be a {\it gauge-variant} 
quantity. Nevertheless, it is somehow accepted that the observation of this
topological index of the (free) helical beams does not contradict the gauge 
principle.
This is probably because there actually exists no difference between the 
canonical OAM and the manifestly gauge-invariant mechanical (or kinetic) 
OAM in the case of free photon or electron beams. 
However, the problem becomes far more intricate,
if one considers recently-investigated helical electron beam propagating
along the direction of a uniform magnetic field \cite{BSVN2012},\cite{SSSLSBN2014}.
In the presence of non-zero magnetic field background, the two OAMs,
the gauge-variant canonical OAM and the gauge-invariant mechanical OAM
become absolutely different quantities. 
Hence, whether the canonical OAM ever corresponds to an observable 
quantity in such nontrivial setups is a fundamental question of physics
with universal meaning.

The main purpose of the present paper is to elucidate the physical meanings
of the canonical OAM and the mechanical OAM as transparently as possible
and then to answer the compatibility question between the gauge principle
and the observability of the {\it gauge-variant} canonical OAM.
First, in sect.2, we briefly remind of the basics of
the Landau problem and explain the motivation of our study.
Next, in sect.3, we point out a novel symmetry of the Landau
eigen-functions which contains the magnetic quantum number $m$. 
With close attention to this symmetry, we analyze in detail 
the structures of the Landau electron's rotational dynamics which 
depends on the quantum number $m$. 
We shall argue that the characteristic of the probability current 
distribution critically depends on the sign of the magnetic quantum 
number $m$ and that it basically explains the three-fold splitting
of the helical electron beam in a magnetic field observed by 
Bliokh et al. \cite{BSVN2012},\cite{SSSLSBN2014}.
In sect.4, with a slight hope of possible observation of the magnetic 
quantum number $m$ of the Landau electron, we analyze the interference
phenomena of two helical electron beams with different magnetic
quantum numbers $m_1$ and $m_2$. By considering varieties of combination
of $m_1$ and $m_2$, we realize that there appear rich structures in
the rotational velocity of the probability densities of the superposition
states. Next in sect.5, we point out a delicate but critical difference
between the nondiffractive Landau beams in a uniform magnetic field and 
the diffractive Laguerre-Gauss (LG) beams in free space. 
We argue that this difference is vitally important for
resolving the compatibility problem between the observability of
the canonical OAM and the gauge principle.   
Finally, in sect.6, we summarize what we have learned from the present study.

\section{Brief reminder of the Landau problem and research motives}
\label{Section2}

The Landau Hamiltonian, which describes the motion of an electron
with charge $- \,e \,(e > 0)$ and mass $m_e$ in the uniform magnetic field, 
is given by \cite{Landau1930},\cite{Landau_Lifschitz1977}
\begin{equation}
 H \ = \ \frac{1}{2 \,m_e} \,\bm{\Pi}^2 \ = \ \frac{1}{2 \,m_e} \,
 \left( \bm{p} \ + \ e \,\bm{A} \right)^2,
\end{equation}
where $\bm{A}$ is the gauge potential, which reproduces the uniform magnetic 
field pointing to the $z$-direction through the standard relation 
$\bm{B} = \nabla \times \bm{A}$. 
(Since only the time-independent gauge symmetry is
considered, only the vector potential $\bm{A}$ is involved in the
following argument.)
Here, $\bm{p}$ and $\bm{\Pi} = \bm{p} \,+ \,e \,\bm{A}$ represent the 
canonical and mechanical momentum operators, respectively.
(We shall use the natural unit $\hbar = c = 1$ throughout the paper.)
Since the Landau Hamiltonian contains {\it gauge-dependent} vector potential
$\bm{A}$, its eigen-functions depend on the choice of gauge.
For our discussion below, it is convenient to work in the so-called symmetric
gauge $\bm{A} = \bm{A}^{(S)} \equiv \frac{1}{2} \,B \,(\,- \,y, \,x)$.
(When we treat the nondiffractive Landau beam propagating along
the direction of the magnetic field, we must consider the 3-dimensional
situation. In this case, since we consider only a uniform magnetic field, 
the $z$-component of the vector potential $A_z$ can be set zero without 
loss of generality.)
As is well-known, the eigen-functions of the Landau Hamiltonian in the 
symmetric gauge are specified by two quantum numbers $n$ and $m$ 
as \cite{LW1999},\cite{FL2000},\cite{VGW2006}
\begin{equation}
 \psi_{n,m} (r, \phi) \ = \ \frac{e^{\,i \,m \,\phi}}{\sqrt{2 \,\pi}} \,\,
 R_{n,m} (r) ,
\end{equation}
where
\begin{equation}
 R_{n,m} (r) \ = \ N_{n,m} \,\,\left( \frac{r^2}{2 \,l^2_B }\right)^{|m|/2} \,
 e^{\,- \,\frac{r^2}{4 \,l^2_B}} \,
 L^{|m|}_{n - \frac{|m|+m}{2}} \,\left( \frac{r^2}{2 \,l_B^2} \right) , 
\end{equation}
with $N_{n,m}$ being the normalization constant given by
\begin{equation}
 N_{n,m} \ = \ (- \,1)^{\,n \,- \,\frac{|m| + m}{2}} \,\,\frac{1}{l_B} \,\,
 \sqrt{\frac{\left(\, n \, - \, \frac{|m| + m}{2} \right) \,!}
 {\left( n \,+ \,\frac{|m| - m}{2} \right) \,!}} .
\end{equation}
Here, $l_B = 1 \,/ \sqrt{e \,B}$ is the so-called the magnetic length in the
Landau problem, and $L^\alpha_n (x)$ is the associated (or generalized) 
Laguerre polynomials.
The quantum number $n$ takes any non-negative integers, i.e.
$n = 0, 1, 2, \cdots$, and is widely called the Landau quantum number, 
while another quantum number $m$ takes any integers subject to the
constraint $m \leq n$ for a given $n$. More specifically,
written above are the simultaneous eigen-functions of the Landau 
Hamiltonian $H$ and the canonical orbital angular momentum (OAM) 
operator $L^{can}_z \equiv (\bm{r} \times \bm{p})_z$, which satisfy the following
eigen-value equations : 
\begin{eqnarray}
 H \,\,\psi_{n,m} (r, \phi) &=& E_n \,\psi_{n,m} (r,\phi), \\
 L^{can}_z \,\psi_{n,m} (r, \phi) &=& m \,\psi_{n,m} (r, \phi).
\end{eqnarray}
Here, the eigen-energies $E_n$ are given by
\begin{equation}
 E_n \ = \ \left( n \ + \ \frac{1}{2} \right) \,\omega_c ,
\end{equation}
with $\omega_c = \frac{e \,B}{m_e}$ being the familiar cyclotron frequency.
We recall that the expectation value of the mechanical OAM operator
$L^{mech}_z \equiv (\bm{r} \times \bm{\Pi})_z$ in the Landau eigen-states are
given by
\begin{equation}
 \langle \psi_{n,m} \,|\,L^{mech}_z \,| \,\psi_{n,m} \rangle
 \ = \ 2 \,n \ + \ 1, 
\end{equation}
which is proportional to the observable Landau energies.
In contrast, it is known that the expectation value of the canonical OAM 
operator, which is nothing but the quantum number $m$ specifying the Landau 
states, does not correspond to any direct observables. Undoubtedly,
non-observable nature of the quantum number $m$ is inseparably 
connected with infinitely many degeneracy of the Landau levels on $m$, as
exemplified by
\begin{eqnarray}
 n \ = \ 0, &\,&  
 m \ = \ 0, \,- \,1, \,- \,2, \,- \,3, \,- \,4, \, \cdots 
 \ \ \ \Rightarrow \ \ \ E_n \ = \ \frac{1}{2} \,\omega_c ,  
 \label{Eq:degeneracy1} \\
 n \ = \ 1, &\,& 
 m \ = \ 1, \, 0, \,- \,1, \,- \,2, \,- \,3,  \,\cdots 
 \ \ \ \ \ \ \Rightarrow \ \ \ E_n \ = \ \frac{3}{2} \,\omega_c , 
 \label{Eq:degeneracy2} \\
 n \ = \ 2, &\,& 
 m \ = \ 2, \, 1, \, 0, \, - \,1, \,- \,2, \,\cdots 
 \ \ \ \ \ \ \ \ \ \Rightarrow \ \ \ E_n \ = \ \frac{5}{2} \,\omega_c . 
 \label{Eq:degeneracy3} \\
 &\,&  \hspace{10mm} \cdots \nonumber
\end{eqnarray}

However, in recent studies of the helical electron beam propagating along the 
direction of a uniform magnetic field,  Bliokh et al. observed a peculiar splitting of
the Landau levels depending on the sign of $m$, i.e. depending on the canonical 
OAM. Does it mean an observation of the quantum number $m$ in the
Landau states ? It seems to us that this is an perplexing observation which needs 
clarification, considering that observation of {\it gauge-variant} canonical OAM 
appears to contradict a fundamental principle of physics called the gauge principle.

Their investigations start from noticing close resemblance between the Landau
states of the electron represented as
\begin{equation}
 \psi^L_{p,m} (r, \phi, z) \ \propto \ 
 \left( \frac{r^2}{l_B^2} \right)^{|m|/2} \,\,L^{|m|}_p \left( \frac{r^2}{2 \,l^2_B} \right)
 \,e^{\,- \,\frac{r^2}{4 \,l^2_B}} \,e^{\,i \,( m \,\phi \ + \ k_z \,z )},
 \label{Eq:Landau_beam}
\end{equation}
with $p \equiv n  -  \frac{|m| + m}{2}$ being the number of nodes in the 
radial part of the wave function, and the (free) electron Laguerre-Gauss (LG)
beam propagating along the $z$-direction \cite{ABSW1992},\cite{APB1999}
\begin{eqnarray}
 \psi^{LG}_{p,m} (r, \phi, z) &\propto& 
 \left( \frac{r^2}{w^2 (z)} \right)^{|m|/2} \,
 L^{|m|}_p \left( \frac{2 \,r^2}{w^2 (z)} \right) \,
 e^{\,- \,\frac{r^2}{w^2 (z)} \ + \ i \,k \,\frac{r^2}{R (z)} } \nonumber \\
 &\,& \hspace{4mm} \times \ \ e^{\,i \,(m \,\phi \ + \ k \,z)} \,
 e^{\,- \, (2 \,p \ + \ |m| \ + \ 1) \,\arctan \,\left( \frac{z}{z_R} \right)} ,
 \label{Eq:LG_beam}
\end{eqnarray}
with the identification $2 \,l_B \leftrightarrow w (z)$. Here, $l_B$ represents
the previously-mentioned magnetic length of the Landau problem, while $w (z)$ stands
for the (weakly) $z$-dependent transverse width of the LG beam. 
The other quantities in the expression of the LG beam can be found in standard 
literatures of laser physics or helical electron beam 
physics \cite{ABSW1992}\nocite{APB1999}\nocite{TT2011}-\cite{BBSN2007}.

A remarkable observation by Bliokh et al. is that the rotation of electrons in a
uniform magnetic field is drastically different from the classical 
cyclotron motion \cite{BSVN2012}.
According to them, instead of rotation with a single cyclotron frequency 
$\omega_c = e \,B /\,m_e$, the Landau electrons, while propagating along 
the direction of the magnetic fields, receive characteristic rotation with three 
different angular velocities, depending on the eigen-value $m$ of the
canonical OAM operator $L^{can}_z = (\bm{r} \times \bm{p})_z$ : 
\begin{equation}
 \langle \omega \rangle \ = \ \left\{ \begin{array}{cc}
 0 \ \ & \ \ (m < 0), \\
 \omega_L \ \ &\ \  (m = 0), \\
 \omega_c \ \ & \ \ \, (m > 0), \\
 \end{array} \right. \label{Eq:three_fold_splitting}
\end{equation}
where $\omega_c$ is the cyclotron frequency, while 
$\omega_L = \omega_c \,/\, 2$ is the Larmor frequency.
In the following analyses, we shall look for a clear physical explanation on the 
above peculiar splitting of the Landau levels with a special intention of 
testing the validity of the gauge principle.

\section{Landau electron's probability and probability current distributions}
\label{Section3}

To answer the questions raised in the previous section, we need a
thorough understanding of the $m$-dependent rotational dynamics of 
the Landau electron.
As argued in our previous letter paper \cite{WKZZ2020},
the following way of looking at the Landau problem is particularly helpful.
That is, as first pointed out by Johnson and Lippmann a long time
ago \cite{DM1966}, the Landau Hamiltonian
$H = \frac{1}{2 \,m_e} \,(\bm{p} + e \,\bm{A})^2$ in the symmetric gauge 
can be expressed as a sum of the two pieces, i.e. the Hamiltonian of the 
familiar 2-dimensional harmonic oscillator and the Zeeman term as
\begin{equation}
 H \ = \ H_{osc} \ + \ H_{Zeeman} ,
\end{equation}
where
\begin{eqnarray}
 H_{osc} \ \ &=& \frac{1}{2 \,m_e} \,(p^2_x + p^2_y) \ + \ 
 \frac{1}{2} \,m_e \,\omega_L^2 \,(x^2 + y^2), \\
 H_{Zeeman} &=& \omega_L \,L^{can}_z .
\end{eqnarray}
Here, $\omega_L$ is the Larmor frequency, while $L^{can}_z$
is just the canonical OAM operator defined by
$L^{can}_z = (\bm{r} \times \bm{p})_z$. The eigen-functions and the
associated eigen-energies of the 2-dimensional harmonic oscillator are
textbook material and they are given as
\begin{equation}
 H_{osc} \,\tilde{\psi}_{p,m} (r, \phi) \ = \ 
 (\,2 \,p + |m| + 1 \,) \,\omega_L \,\tilde{\psi}_{p,m} (r, \phi),
\end{equation}
where
\begin{equation}
 \tilde{\psi}_{p,m} (r,\phi) \ = \ \frac{e^{\,i \,m \,\phi}}{\sqrt{2 \,\pi}} \,\,
 \tilde{R}_{p,m} (r) , \label{Eq:psi_tilde}
\end{equation}
with
\begin{equation}
 \tilde{R}_{p, m} (r) \ = \ (- \,1)^p \,\,\frac{1}{b} \,
 \sqrt{\frac{2 \,p !}{(p + |m|) \,!}} \,e^{\,- \,\frac{r^2}{2 \,b^2}} \,
 \left( \frac{r^2}{b^2} \right)^{|m| / 2} \,
 L^{|m|}_p \left( \frac{r^2}{b^2} \right) , \label{Eq:radial_wf}
\end{equation}
and with $b^2 = 1 \,/\,(m_e \,\omega_L) = 2 \,/\,(e \,B) = 2 \,l^2_B$.
In the above equations, $p$ stands for the
number of radial nodes, which takes zero or any positive integers.
On the other hand, $m$ represents the azimuthal 
or magnetic quantum number, which is the eigen-value of the canonical OAM 
operator $L^{can}_z = - \,i \,\frac{\partial}{\partial \phi}$ : 
\begin{equation}
 L^{can}_z \,\tilde{\psi}_{p, m} (r, \phi) \ = \ m \,\tilde{\psi}_{p, m} (r, \phi) ,
\end{equation}
with $m$ taking any (positive, zero, or negative) integers. Since $\tilde{\psi}_{p, m}$
are simultaneous eigen-functions of $H_{osc}$ and $H_{Zeeman}$, it immediately
follows that they are also the eigen-functions of the whole Landau Hamiltonian : 
\begin{equation}
 H \,\tilde{\psi}_{p, m} (r, \,\phi) \ = \ E \,\tilde{\psi}_{p, m} (r, \,\phi) ,
\end{equation}
with the corresponding eigen-energies,
\begin{equation}
 E \ = \ E_{osc} \ + \ E_{Zeeman} \ = \ 
 \left[\, (2 \,p + |m| + 1) \ + \ m \,\right] \,\omega_L .
\end{equation}
Here, $E_{osc}$ and $E_{Zeeman}$ respectively stand for the energy of the 
2-dimensional harmonic oscillator and that of the Zeeman term.

Already at this stage, we notice a remarkable fact. If the sign of the magnetic 
quantum number $m$ is negative, the energy of the Landau state becomes
\begin{equation}
 E \,(m < 0) \ = \ ( 2 \,p \ + \ 1) \,\omega_L , 
\end{equation}
irrespectively of the absolute magnitude of $m$. 
This explains the reason why the Landau states with $m < 0$ are infinitely
degenerate as illustrated by 
Eqs.(\ref{Eq:degeneracy1})-(\ref{Eq:degeneracy3}).  
What is worthy of special mention here is how the cancellation of these 
$m$-dependent terms happens. 
Note that the $m$-dependent part of the energy of the 2-dimensional
harmonic oscillator depends only on the absolute value of $m$.
This is related to the time-reversal symmetry of the 2-dimensional harmonic
oscillator Hamiltonian. Physically, it means
that the system energy is independent on the
direction of the electron's rotational motion specified by the sign of $m$.
On the other hand, since the action of the magnetic field breaks the
time-reversal symmetry, the Zeeman energy depends on the sign of $m$
or on the direction of rotation.
The infinite degeneracy of the Landau states with $m < 0$ can therefore be
understood as an interplay of the two different rotational dynamics of
the Landau electron. In any case, what is disclosed by the above
consideration is a very strange property of the Landau states with 
negative $m$ and a simple explanation about its origin.  
(Although it is left out in the present paper, the singular nature of the 
Landau state with 
$m < 0$ has another more intuitive explanation based on the quantum 
{\it guiding center} concept in the Landau problem \cite{WKZZ2020}.)

In the standard representation of the Landau eigen-states,
it is customary to introduce a new 
quantum number $n$ defined by
\begin{equation}
 n \ \equiv \ p \ + \ \frac{|m| + m}{2} .
\end{equation}
This quantum number takes zero or any positive integer and is called the 
Landau quantum number. (Note that, from the above 
relation between $n$, $p$, and $m$, the quantum number $m$ must satisfy the
inequality $m \leq n$, that is, the maximum value of $m$ is $n$ for a given $n$,
as already exemplified by eqs.(7)-(8).)
Accordingly, the eigen-functions of the Landau Hamiltonian are standardly
expressed in terms of $n$ and $m$ instead of $p$ and $m$, which motivates to
define new wave functions $\psi_{n, m} (r)$ by 
$\psi_{n, m} (r, \phi) \equiv \tilde{\psi}_{p, m} (r, \phi)$.
As a consequence, the eigen-energies of the Landau Hamiltonian depend only 
on the quantum number $n$ as
$H \, \psi_{n, m} (r, \phi) \ = \ (2 \,n + 1) \,\omega_L \,\,\psi_{n, m} (r, \phi)$.

These are basically known stories, but the fact that the Landau eigen-states 
are also the eigen-states of the 2-dimensional harmonic oscillator makes 
us aware of an remarkable symmetry of the Landau eigen-functions
as first pointed out in \cite{WKZZ2020}.
First, as can be explicitly convinced from Eq.(\ref{Eq:radial_wf}),
the radial wave functions $\tilde{R}_{p, m} (r)$ of the
2-dimensional harmonic oscillator have a very simple symmetry : 
\begin{equation}
 \tilde{R}_{p, - \,m} (r) \ = \ \tilde{R}_{p, m} (r),
\end{equation}
i.e. the symmetry under the reverse of the magnetic quantum number $m$.
It is clear that this symmetry also originates from the time-reversal invariance
of the Hamiltonian of the 2-dimensional harmonic oscillator. Very interestingly, if
this symmetry of $\tilde{R}_{p, m} (r)$ is translated into the symmetry of
the standard form of radial wave functions in the Landau problem, defined
by $R_{n, m} (r) \equiv \tilde{R}_{p, m} (r)$ with $b = \sqrt{2} \,l_B$, we are led to 
a highly nontrivial relation represented as
\begin{equation}
 R_{n-m, - \,m} (r) \ = \ R_{n, m} (r) .
\end{equation}

To understand how surprising this symmetry relation is, let us, as an
example, consider a special case where $n = m = 20$. In this particular case, 
one has the relation : 
\begin{equation}
 R_{0, - \,20} (r) \ = \ R_{20, 20} (r) .
\end{equation}
In view of Eq.(\ref{Eq:psi_tilde}), this relation dictates that the probability density 
$|\psi_{0, -\,20} (r,\phi)|^2$
of the state with $n = 0$ and $m = - \,20$ is exactly the same as the probability density
$|\psi_{20,20} (r,\phi)|^2$ of the state with $n = 20$ and $m = 20$.
Note however that the eigen-energies of these two states are totally different as
seen from
\begin{eqnarray}
 E (n = 0, m = - \,20) &=& (\, 2 \times 0 \ + \ 1 \,) \,\omega_L \ = \ \ \omega_L, \\
 E (n = 20, m = + \,20) &=& (\, 2 \times 20 \ + \ 1 \,) \,\omega_L \ = \ 41 \,\omega_L .
\end{eqnarray}
We thus observe that, while these two states have exactly the same probability
densities, they have totally different eigen-energies. As we shall see below, the
cause of this peculiar observation can be traced back to the fact that, 
although the probability densities of these two states are exactly the same, 
they have totally different probability current distributions.
The point is that, under the presence of the external magnetic field, the
internal electric current carried by the electron interacts with this magnetic field,
and this interaction also contributes to the energy of the system. This means that,
if the two states in question have different probability current distributions,
they generally have different energies even if they have the identical
probability distributions.

To convince the statements above in more concrete manner, we first recall that the 
total probability current density of the electron is given by the following familiar 
expression : 
\begin{equation}
 \bm{j} \ = \ \frac{1}{m_e} \,\mbox{Im} \,(\,\psi^* \,\bm{D} \,\psi ),
\end{equation}
with $\bm{D} = \nabla + i \,e \,\bm{A}$ being the so-called covariant derivative.
Note that the net current shown above is gauge-invariant.
(Here and hereafter, we use the word {\it gauge-invariant} following the 
custom of prevailing literatures. More precisely, however, we should use the
word {\it gauge-covariant}. In fact, what is gauge-invariant is the
expectation value of the above current operator.)
Obviously, this total current consists of two pieces as
\begin{equation}
 \bm{j} \ = \ \bm{j}^{\,can} \ + \ \bm{j}^{\,gauge},
\end{equation}
with
\begin{equation}
 \bm{j}^{\,can} \ = \ \frac{1}{m_e} \,\mbox{Im} \,(\,\psi^* \,\nabla \,\psi),
 \hspace{6mm} \bm{j}^{\,gauge} \ = \ \frac{1}{m_e} \,\psi^* \,e \,\bm{A} \,\psi ,
\end{equation}
which we hereafter call the canonical current and the gauge current,
respectively. (They are sometimes called the {\it paramagnetic}
current and the {\it diamagnetic} current in the field of electronic
physical properties \cite{GFS2015}.) Widely-accepted viewpoint is 
that the canonical current and the 
gauge current are not separately gauge-invariant. 
Only the sum of them is gauge-invariant.

In the Landau states described by the eigen-functions 
(\ref{Eq:psi_tilde}) and (\ref{Eq:radial_wf}), both
parts of current have only the azimuthal components as 
$\bm{j}^{\,can} = j^{\,can}_\phi \,\bm{e}_\phi$
and $\bm{j}^{\,gauge} = j^{\,gauge}_\phi \,\bm{e}_\phi$, where
\begin{equation}
 j^{\,can}_\phi \ = \ \frac{1}{m_e} \,\frac{m}{r} \,\rho (r), \hspace{6mm}
 j^{\,gauge}_\phi \ = \ \frac{1}{m_e} \,\frac{r}{2 \,l^2_B} \,\rho (r) ,
 \label{Eq:jcan_jgauge}
\end{equation}
with $\rho (r) = | \psi_{n, m} |^2$ being the electron probability density.
Here, the probability density is normalized as
\begin{equation}
 \int_0^{2 \,\pi} \,d \phi \,\int_0^\infty \,d r \,r \,\,\rho (r)
 \ = \ 2 \,\pi \,\int_0^\infty \,d r \,r \,\,\rho (r) \ = \ 1.
\end{equation}
Note that, due to the axial symmetry of the Landau eigen-states in
the symmetric gauge, $\rho$ is a function of the radial coordinate $r$ alone.
For convenience, we also introduce the dimensionless probability
density $\tilde{\rho} (R)$ and the dimensionless current density
$\tilde{j}_\phi (R)$ by $\tilde{\rho} (R) \equiv l^2_B \,\rho (r)$ and
$\tilde{j}_\phi (R) \equiv m_e \,l^3_B \,j_\phi (r)$ with 
$R \equiv r / l_B$ being the dimensionless radial coordinate. 

\begin{figure}[ht]
\begin{center}
\includegraphics[width=8cm]{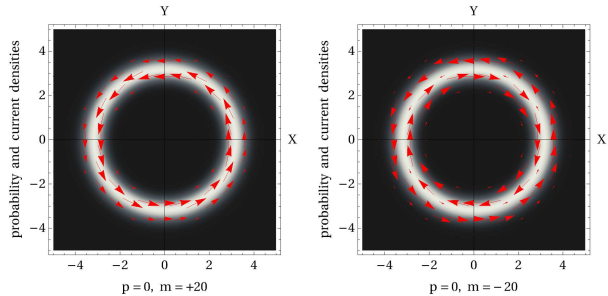}
\vspace{-2mm}
\caption{The grey scale images on the left and right panels of 
this figue respectively show 
the 2-dimensional plots of the probability distributions 
corresponding to the Landau state with $p = 0, m = 20$ and that with 
$p = 0, m = - \,20$ in units of dimensionless coordinates 
$X = x \,/\, l_B$ and $Y = y \,/\,l_B$.
Shown by arrows (red in color) in both panels are
the dimensionless (total) probability current distributions corresponding to the
Landau state with $p = 0, m = 20$ and that with $p = 0, m = - \,20$.}
\label{Fig:Prob_Current_total_nd0_p20_m20_BB} 
\end{center}
\end{figure}

In Fig.\ref{Fig:Prob_Current_total_nd0_p20_m20_BB}, 
we compare the probability distributions 
as well as the probability current distribution of the
Landau state with $p = 0, m = 20$ \, (or equivalently $n=20, m = 20$) 
and those with $p = 0, m = - \,20$ \,
(or $n = 0, m = - \,20$).
The greyscale images shown in the left and right panels of this figure
are respectively the 2-dimensional density plots of the probability distributions corresponding
to the Landau state with $p = 0, m = 20$ and that with $p = 0, m = - \,20$ in the 
units of dimensionless coordinates $X = x \,/\, l_B$ and $Y = y \,/\,l_B$.
One can confirm that the probability distribution of the $p = 0, m = 20$
state and that of the $p = 0, m = - \,20$ state perfectly coincide
with each other.
Shown by arrows (red in color) in the same figures are the dimensionless 
probability current densities corresponding to
the Landau state with $p = 0, m = 20$ and that with $p = 0, m = -\,20$.
Taking a closer look at these figures, one sees
that the current distributions of these two states are drastically different.
For the state with $p = 0, m = 20$, the flow of probability current is 
counter-clock wise in the whole region in which the magnitude of
the probability density is appreciable (the brighter region in color).
On the other hand, for the state with $p = 0, m = - \,20$, the flow of
current is counter-clock-wise in the outer part of the high
probability density region, while it is clock-wise in the inner part.

\begin{figure}[ht]
\begin{center}
\includegraphics[width=8cm]{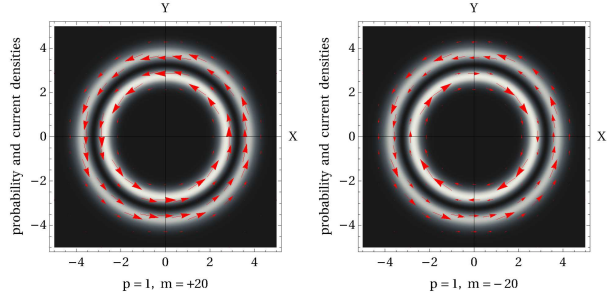}
\caption{The same as Fig.\ref{Fig:Prob_Current_total_nd0_p20_m20_BB} except 
that the left panel correspond 
to the Landau state with $p = 1, m = 20$, while the right panel to the 
Landau state with $p = 1, m = - \,20$.}
\label{Fig:Prob_Current_total_nd1_p20_m20_BB} 
\end{center}
\end{figure}

The above remarkable difference in the behavior of the probability currents 
for the $m > 0$ state and for the $m < 0$ state would even more clearly 
be seen if one looks into the Landau states with one radial node.
Shown by gray scale images in the left and right panels of 
Fig.\ref{Fig:Prob_Current_total_nd1_p20_m20_BB}
are the 2-dimensional plots of the probability densities corresponding to 
the state with $p = 1, m = 20$ and with $p = 1, m = - \,20$, respectively.
Due to the existence of a node of the radial wave function, the
probability densities of these two states show double-ring structure.
Still, the probability densities of these two states are confirmed to be exactly 
the same. 
On the other hand, shown by arrows (red in collor) in the same figures
are the (net) probability current distributions corresponding to
the two Laudau states with $p = 1, m = 20$ and $p = 1, m = -\,20$.
For the state with $p = 1, m = 20$, the flow of probability current shown by 
arrows (red in color) is counter-clock wise in the outer ring as well as 
in the inner ring.
In contrast, for the state with $p = 1, m = - \,20$, the flow of current is 
counter-clock-wise in the outer ring, while it is clock-wise in 
the inner ring.  

\begin{figure}[ht]
\begin{center}
\includegraphics[width=11cm]{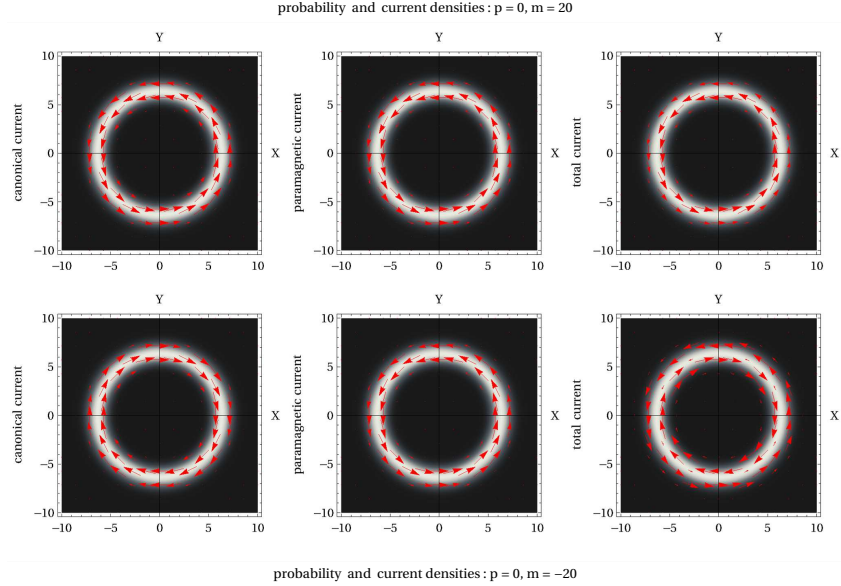}
\caption{Shown in the three figures on the upper and lower panels 
are the probability density and the current densities in unit of $X$ and $Y$.
The upper panel corresponds to the Landau state
with $p = 0$ and $m = + \,20$, while the lower panel to that
with $p = 0$ and $m = - \,20$. 
The arrows (red in color) in the left, middle, and
right figures in both panels represents the canonical current, the gauge current, 
and the total current, respectively. }
\label{Fig:Dens_Current_separate_nd0_p20_m20_BB} 
\end{center}
\end{figure}

A natural question is then how we can understand the above-mentioned 
characteristic difference in the structures of the current distributions
for the $m > 0$ state and for the $m < 0$ state.
As discussed in the paper \cite{BSVN2012} by Bliokh et al., 
it can be understood as an interplay of the
canonical and gauge parts of the probability current.
In Fig.\ref{Fig:Dens_Current_separate_nd0_p20_m20_BB} 
we compare the probability densities and the probability current densities 
in the 2-dimensional $(X, Y)$ plane for the Landau state with $p = 0, m = 20$
(upper panel) and that with $p = 0, m = - \,20$ (lower panel).
The upper panel corresponds to the state with $p = 0, m = 20$ or equivalently to
the state with $n = 20, m = 20$, while the lower panel
to the state with $p = 0, m = - \,20$ or equivalently that with $n = 0, m = - \,20$. 
One can again confirm that the
probability density of the state with $n = 20, m = 20$ shown in 
the upper panel and that with $n = 0, m = - \,20$ shown in the lower panel
perfectly coincide in spite of the fact that their eigen-energies are totally 
different as convinced from their different Landau quantum numbers.
From the behavior of the canonical and the gauge current distributions, 
one can easily understand the reason why the behavior of the probability 
net currents for these two states are remarkably different.
Since $m > 0$ for the state with $p = 0, m = 20$, both of the
canonical current $j^{can}_\phi$ and the gauge current $j^{gauge}_\phi$
are positive, which means that both of the canonical current and the
gauge current are circulating in a counter-clock-wise direction.
Accordingly, total current is also flowing counter-clock-wise.
On the other hand, since $m < 0$ for the state with $p = 0, m = - \,20$,
the canonical current is flowing clock-wise, whereas the gauge current
is flowing counter-clock-wise. An important fact here is that, 
because of different radial dependencies
of the canonical and gauge currents given as 
$j^{can}_\phi (r) \propto \frac{1}{r} \,\rho (r)$ and
$j^{gauge}_\phi (r) \propto r \,\rho (r)$ (see Eq.(\ref{Eq:jcan_jgauge})), the
flow of the total current shows highly nontrivial behavior as illustrated 
in the rightmost figure of the lower panel of this figure. 
Namely, the flow of the net current for the state with $p = 0, m = - \,20$
is counter-clock-wise in the outer part of the high probability density
region, whereas it is clock-wise in the inner part of the high
probability density region.  

\begin{figure}[ht]
\begin{center}
\includegraphics[width=11cm]{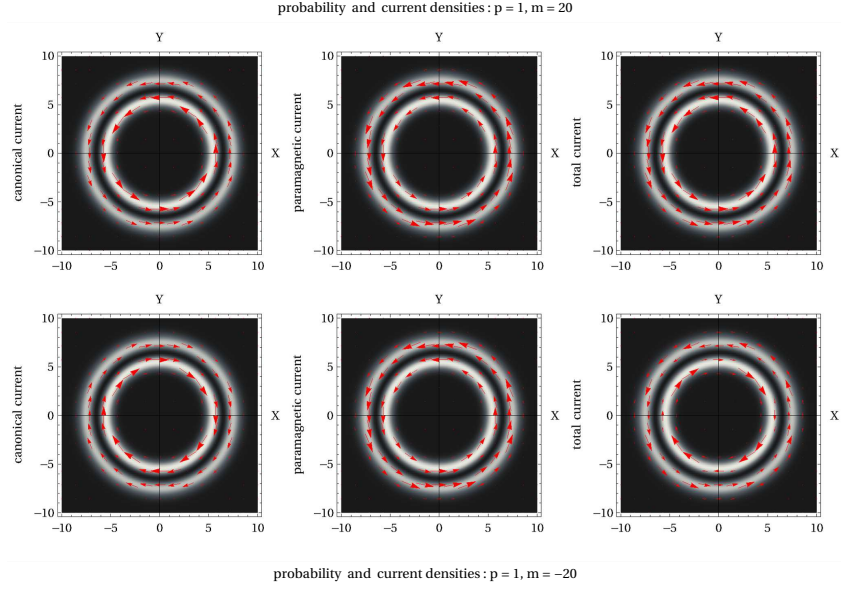}
\caption{The same as Fig.\ref{Fig:Dens_Current_separate_nd0_p20_m20_BB} 
except that all the figures correspond to the
the Landau state with $p = 1, m = \pm \,20$, which has one node in the radial
wave function.}
\label{Fig:Dens_Current_separate_nd1_p20_m20_BB} 
\end{center}
\end{figure}

For the sake of completeness, we show in 
Fig.\ref{Fig:Dens_Current_separate_nd1_p20_m20_BB}
the probability densities and the separate contributions of the
canonical and gauge currents to the probability current densities 
for the state with $p = 1, m = 20$ and that with $p = 1, m = - \,20$.
What we can learn from these figures are basically the same as learned 
from Fig.\ref{Fig:Dens_Current_separate_nd0_p20_m20_BB}  
for the $ p =0$ case, but 
the double-ring structure coming from a node of the radial wave functions
makes it easier to see the interplay between the canonical and gauge 
currents.

So far, we have seen that the interplay of the canonical current and the
gauge current causes a drastic difference in the behavior of the
probability current of the electron, which critically depends on the sign
of the quantum number $m$ in the Landau states.
In fact, it is the origin of the highly nontrivial rotational dynamics of
the Landau electron pointed out by Bliokh et al.
To understand this, it is convenient to introduce electron's angular
velocity distribution $\omega (r)$ related to the azimuthal component of
the probability current density as \cite{LW1999},\cite{SSSLSBN2014}
\begin{equation}
 \omega (r) \ = \ \frac{j_\phi (r)}{r} .
\end{equation}
Naturally, this quantity is also expressed as a sum of the contributions
of the canonical current and the gauge current as
\begin{equation}
 \omega (r) \ = \ \frac{1}{r} \,\left( 
 j^{\,can}_\phi (r) \ + \ j^{\,gauge}_\phi (r) \right)
 \ = \ \frac{1}{m_e} \,\frac{m}{r^2} \,\rho (r) \ + \ 
 \frac{1}{m_e} \,\frac{1}{2 \,l^2_B} \,\rho (r).
\end{equation}
The average angular velocity $\bar{\omega}$ is obtained by integrating
$\omega(r)$ over the whole $xy$-plane as
\begin{equation}
 \bar{\omega} \ = \ 2 \,\pi \,\int_0^\infty \,
 \omega (r) \,r \,d r
\end{equation}
Using the relation
$2 \,\pi \, \int \,\frac{\rho (r)}{r^2} \,r \, dr = 1 \,/\,(2 \,l^2_B \,|m| )$, one 
thus finds that
\begin{equation}
 \bar{\omega} \ = \ \omega_L \,\left(
 \frac{m}{|m|} \ + \ 1 \right) , \label{Eq:angular_velocity}
\end{equation}
which confirms the 3-fold splitting of the Landau electron's rotational
velocity depending on the sign of the magnetic quantum number $m$
already shown in Eq.(\ref{Eq:three_fold_splitting}).
We point out that this relation was already written down in the paper
by Li and Wang \cite{LW1999}, although its practical importance became clear only
after the proposal of using the helical electron beams 
by Bliokh et al. \cite{BSVN2012},\cite{SSSLSBN2014}.

Since an important aim of our present study is to clarify
the physical meanings and the differences of the canonical and mechanical
OAMs of the electron or of any charged particles especially in relation
with their observability or nonobservability, we are also interested
in the angular momentum density $\bm{l}$, 
which is related to the probability current density $\bm{j}$ by
\begin{equation}
 \bm{l} \ = \ m_e \,\bm{r} \times \bm{j} .
\end{equation}
Since the probability current density is made up of the canonical part and the
gauge part, the total angular momentum density also consists of the
two pieces as
\begin{equation}
 \bm{l} \ = \ \bm{l}^{\,can} \ + \ \bm{l}^{\,gauge},
\end{equation}
with
\begin{equation}
 \bm{l}^{\,can} \ = \ m_e \,\bm{r} \times \bm{j}^{\,can}, \hspace{8mm}
 \bm{l}^{\,gauge} \ = \ m_e \,\bm{r} \times \bm{j}^{\,gauge} .
\end{equation}
The total orbital angular momentum (OAM) density $\bm{l}$ is called by 
various names such as the kinetic OAM, mechanical OAM, or the dynamical 
OAM \cite{Book_Sakurai1995}.
In the present paper, we shall mostly use the terminology, the mechanical OAM.
As is widely-believed, this mechanical OAM is a gauge-invariant quantity,
whereas the canonical OAM as well as the gauge part of the OAM are
not separately gauge-invariant.  
We recall that the Landau eigen-states in the symmetric gauge is
the simultaneous eigen-states of the Landau Hamiltonian and the
canonical OAM operator. However, one should keep in mind the fact that
the Landau eigen-states in other gauges like the so-called Landau gauges
are not the eigen-states of the canonical OAM operator \cite{WKZZ2018}.

For the Landau eigen-states in the symmetric gauge,
both of the canonical OAM and the gauge part of the OAM have only the
$z$-component and take the following form
\begin{equation}
 l^{\,mech}_z \ = \ l^{\,can}_z \ + \ l^{\,gauge}_z ,
\end{equation}
with
\begin{equation}
 l^{\,can}_z \ = \ m \,\rho (r), \hspace{8mm} 
 l^{\,gauge}_z \ = \ \frac{r^2}{2 \,l^2_B} \,\,\rho (r) . \label{Eq:lcan_lgauge}
\end{equation}
The spatial integrals of these densities coincide with the expectation values 
of the relevant OAM operators $L^{mech}_z, L^{can}_z$ and $L^{gauge}_z$
in the Landau eigen-states. 
It is easy to show that the expectation 
values of these OAM operators are
given by 
\begin{equation}
 \langle L^{can}_z \rangle \ = \ m, \hspace{8mm}
 \langle L^{gauge}_z \rangle \ = \ 2 \,n \ + \ 1 \ - \ m,
\end{equation}
which in turn gives 
\begin{equation}
 \langle L^{mech}_z \rangle \ = \ \langle L^{can}_z \rangle \ + \ 
 \langle L^{gauge}_z \rangle \ = \ 2 \,n \ + \ 1 .
\end{equation}
One observes that, in the expectation value of the mechanical OAM
operator, the quantum number $m$ appearing in the expectation
value of the canonical OAM is exactly canceled by the 
$m$-dependent part appearing in the
gauge part of the OAM. As a consequence, the expectation value of
the mechanical OAM depends only on the Landau quantum number $n$
and is proportional to the observable Landau energy, which is a well-known
fact. (See, for example, \cite{WKZ2018}.)

Before ending this section, we recall that Johnson and Lippmann also
pointed out that the magnetic moment $\mu_z$ of the Landau system
is obtained from the relation \cite{JL1949}
\begin{equation}
 \mu_z \ = \ \frac{\partial H}{\partial B} ,
\end{equation}
where $H = \frac{1}{2 \,m_e} \,(\Pi^2_x + \Pi^2_y )$ is the Landau
Hamiltonian in the symmetric gauge.
Using the relation
\begin{eqnarray}
 \frac{\partial \Pi_x}{\partial B} &=& \frac{\partial}{\partial B} \,
 \left( - \,\frac{1}{2} \,e \,B \,y \right) \ = \ - \,\frac{1}{2} \,e \,y , \\
 \frac{\partial \Pi_y}{\partial B} &=& \frac{\partial}{\partial B} \,
 \left( + \,\frac{1}{2} \,e \,B \,x \right) \ = \ + \,\frac{1}{2} \,e \,x ,
\end{eqnarray}
together with the commutation relation
\begin{equation}
 [\,x, \Pi_y ] \ = \ 0, \hspace{8mm} [\,y, \Pi_x ] \ = \ 0,
\end{equation}
we are led to a simple relation
\begin{equation}
 \mu_z \ = \ \frac{1}{m_e} \,\left( \Pi_x \,\frac{\partial \Pi_x}{\partial B} \ + \ 
 \Pi_y \,\frac{\partial \Pi_y}{\partial B} \right) \ = \ 
 \frac{e}{2 \,m_e} \,\left( x \,\Pi_y \ - \ y \,\Pi_x \right) \ = \ 
 \frac{e}{2 \,m_e} \,L^{mech}_z .
\end{equation}
This relation just reconfirms that the magnetic moment of the Landau 
system originates from the cyclotron motion of the electron. 
Taking the expectation value in the Landau state, it gives
\begin{equation}
 \langle \bm{B} \cdot \bm{\mu} \rangle \ = \ \omega_L \,
 \langle L^{mech}_z \rangle , \label{Eq:Bmyu}
\end{equation}
where we have used the relation $\omega_L = e \,B \,/\,(2 \,m_e)$.
Remember here the fact that the Landau Hamiltonian is
given as a sum of the 2-dimensional harmonic oscillator Hamiltonian
$H_{osc}$ and the Zeeman term $H_{Zeeman} = \omega_L \,L^{can}_z$.
Furthermore, there is a simple relation resulting
from the Virial theorem that relates the expectation value of the kinetic 
term and that of the potential term
in the 2-dimensional harmonic oscillator Hamiltonian : 
\begin{equation}
 \left\langle \, \frac{1}{2 \,m_e} \,(p^2_x + p^2_y) \,\right\rangle \ = \ 
 \left\langle \,\frac{1}{2} \,m_e \,\omega^2_L \,(x^2 + y^2) \,\right\rangle .
\end{equation}
By using this relation, the expectation value of the Landau Hamiltonian
can be written as
\begin{equation}
 \langle H \rangle \ = \ \omega_L \,\langle \,L^{can}_z \ + \ 
 m_e \,\omega_L \,r^2 \, \rangle .
\end{equation}
Here, for the eigen-states in the symmetric gauge, it holds that
\begin{equation}
 \langle \,m_e \,\omega_L \,r^2 \,\rangle \ = \ 
 \langle L^{gauge}_z \rangle .
\end{equation}
This therefore gives the relation
\begin{equation}
 \langle H \rangle \ = \ \omega_L \,\,\langle L^{can}_z \ + \ 
 L^{gauge}_z \rangle \ = \ 
 \omega_L \,\,\langle L^{mech}_z \rangle . \label{Eq:H_Lmech}
\end{equation}
Comparing (\ref{Eq:Bmyu}) and (\ref{Eq:H_Lmech}), we are then led to a 
remarkable relation 
\begin{equation}
 \langle \bm{B} \cdot \bm{\mu} \rangle \ = \ \langle H \rangle ,
\end{equation}
with
\begin{equation}
 \bm{\mu} \ = \ \frac{e}{2 \,m_e} \, \bm{L}^{can} \ + \ 
 \frac{e}{2 \,m_e} \,\bm{L}^{gauge} \ \equiv \ 
 \bm{\mu}^{\,can} \ + \ \bm{\mu}^{\,gauge}
\end{equation}
This is also an expected relation, since the Landau energy can be thought
to come from the interaction between the external magnetic field and the 
magnetic moment originating from the cyclotron motion of the electron.
(Note that the spin degrees of freedom of the electron is totally 
neglected in the present paper.)
One might call the above energy a generalized Zeeman energy
in the sense that the Zeeman term $\omega_L \,L^{can}_z$ is only
a part of this total energy of Zeeman type. 
An important fact learned from the above consideration is that
the interaction between the external magnetic field and
the electron's current always appears in a {\it single combination} of the 
canonical and gauge currents.
This implies that each part of the total current cannot be separately
measured. According to common wisdom, one would say that 
this is not unrelated to the 
fact that only the net current or the mechanical OAM is gauge-invariant, while
the canonical part and the gauge part are not separately gauge-invariant.

Still, we should be ready for the following objection. Namely,
someone might say that the arguments above is primarily based on the 
{\it particle aspect} of the electron even though its theoretical foundation
is quantum mechanics. 
In fact, in a recent paper \cite{GSFB2014}, Greenshields et al. argue that the 
canonical and mechanical
momenta (and also the corresponding OAMs) are associated with {\it wavelike}
and {\it particlelike} properties, respectively. This appears to suggest a possibility
that the canonical momentum and/or the canonical OAM might be observed 
by making use of a wavelike property of the electron beam, say, through some 
interference phenomenon of electron wave functions.
In the subsequent sections, we shall discuss this very delicate issue 
with minute attention.

\section{Interference of helical electron beams with two different
orbital angular momenta}
\label{Section4}  

In the typical Landau problem, the motion of the electron along the
$z$-direction is neglected or simply treated as a free plane-wave state,
so that the problem essentially reduces to a two-dimensional one.
As argued in the previous section, the 3-fold splitting of the
electron's average angular velocity $\langle \omega (r) \rangle$ can 
essentially be understood as an interplay of the canonical and gauge-field 
parts of the electron's probability current in the 2-dimensional transverse plane.
In the simple 2-dimensional setup, however, there is no way to observe
this splitting of the angular velocity of the electron.
The wisdom of Bliokh et al. for observing this novel splitting is to
consider the Landau vortex mode represented by (\ref{Eq:Landau_beam})
and their superpositions generated in a system with a fixed electron
energy $E$ and free propagation along the 
$z$-direction \cite{BSVN2012},\cite{SSSLSBN2014}.
Such beams are prepared in the following way.
Let us first introduce a longitudinal Larmor length $z_m$
determined by the Larmor frequency $\omega_L$ and the electron
velocity $v = \sqrt{2 \,E / m_e}$ as 
\begin{equation}
 z_m \ = \ \frac{v}{\omega_L} .
\end{equation}
Note that, by using the transverse magnetic length $w_m = 2 \,/ \sqrt{e \,B} = 2 \,l_B$,
this longitudinal scale can also be expressed as 
\begin{equation}
 z_m \ = \ \frac{2 \,\sqrt{2 \,E \,m_e}}{e \,B} \ = \ \sqrt{\frac{E}{\omega_L}} \,\,w_m .
\end{equation}
Here, different modes will have different wave number $k_z$ that satisfies
the dispersion relation
\begin{equation}
 E \ = \ E_\parallel \ + \ E_\perp ,
\end{equation}
where $E_\parallel = k^2_z / (2 \,m_e)$ represents the energy 
corresponding to the electron's motion along the $z$-direction, whereas
\begin{equation}
 E_\perp \ = \ E_Z \ + \ E_G, \label{Eq:E_Z_G}
\end{equation}
stands for the energy corresponding to the electron's motion in the transverse plane,
i.e. the plane perpendicular to the $z$-axis. 
Here, the first part $E_Z = m \,\omega_L$ corresponds to
the Zeeman energy, while the second part $E_G$ corresponds to the eigen-energy
of the 2-dimensional harmonic oscillator discussed in sect.3.
In \cite{BSVN2012}, this second part was called the 
Gouy term because it can be related to the Gouy phase appearing in the
diffractive LG beam represented by (\ref{Eq:LG_beam}).
Under the condition that the paraxial conditions $E_\perp \ll E$ and $w_m \ll z_m$
are satisfied, the wave number $k_z$ can approximately be written as
\begin{equation}
 k_z \ \simeq \ k \ + \ \Delta k_z ,
\end{equation}
where $k = \sqrt{2 \,E \,m_e}$, and
\begin{equation}
 \Delta k_z (p, m) \ = \ - \,\left[ m \ + \ (\,2 \,p + |m| + 1) \right] \,/\,z_m .
\end{equation}
This gives
\begin{equation}
 e^{\,i \,k_z \,z} \ \simeq \ e^{\,i \,k \,z} \,\,e^{\,i \,\Delta k_z \,z} 
 \ = \ e^{\,i \,k \,z} \,\,e^{\,i \,\Phi_{LZG}} ,
\end{equation}
with
\begin{equation}
 \Phi_{LZG} \ = \ - \,\left[ m \ + \ ( 2 \,p  + |m| + 1) \right] \,z \,/\,z_m .
 \label{Eq:LZG_phase}
\end{equation}
This additional phase $\Phi_{LZG}$ was called the Landau-Zeeman-Gouy 
phase in \cite{BSVN2012}. We emphasize the fact that 
this extra phase depends on the two quantum numbers
$p$ and $m$ only through the combination $m + ( 2\,p + |m| +1)$,
which is equal to $2 \,n + 1$ if we use the Landau quantum number
defined by $n \equiv p + (|m| + m) / 2$.

In sum, the above consideration reveals that, while propagating along
the $z$-direction, the magnetic field correction to the longitudinal
wave number $k_z$ yields an additional phase
to the wave functions. After this correction is taken into account, 
the superposed nondiffractive Landau beam becomes
\begin{eqnarray}
 \psi_{p, m} (r, \phi, z) &\simeq& \frac{N_{p, m}}{\sqrt{2 \,\pi}} \,
 \left( \frac{2 \,r^2}{w^2_m} \right)^{|m| \,/\,2} \,\,
 L^{|m|}_p \left( \frac{2 \,r^2}{w^2_m} \right) \,\,
 e^{\,- \,\frac{r^2}{w^2_m}} \,e^{\,i \,k \,z} \,\,
 e^{\,i \,( m \,\phi \ + \ \Delta k_z (p, m) \,z )} ,
\end{eqnarray}
with $N_{p,m}$ being the normalization constant of the Landau state
given by
\begin{equation}
 N_{p, m} \ = \ \frac{2}{w_m} \,\sqrt{\frac{p \,!}{(p + |m|) \,!}},
\end{equation}
This means that the electron wave function acquires an extra phase
$\Phi_{LGZ} = \Delta k_z (p, m) \,z$, which changes as a function of the 
propagation distance $z$ along the directions of the 
magnetic field. However, this $z$-dependent phase change of the wave 
function cannot be detected straightforwardly by observing the beam intensity.
In fact, the probability density of the electron beam is given by
\begin{equation}
 \rho \ = \ | \psi_{p, m} |^2 \ = \ \frac{N^2_{p, m}}{2 \,\pi} \,
 \left( \frac{2 \,r^2}{w^2_m} \right)^{|m|} \,
 \left[ L^{|m|}_p \left( \frac{2 \,r^2}{w^2_m} \right)\right]^2 \,
 e^{\,- \,\frac{2 \,r^2}{w^2_m}} ,
\end{equation}
which is axially-symmetric with respect to the $z$-axis and is 
independent of the coordinate $z$. This is the very reason why the 
authors of \cite{SSSLSBN2014} had to invent a clever experiment in which 
half of the beam is obstructed to stop with an opaque knife edge stop and 
the spatial rotation of the visible part of the beam is traced by moving the knife edge
along the beam direction, in order to verify the $m$-dependent
splitting of the electron helical beam.

Observation of the phase rotation of the wave function would however be 
possible if one constructs a superposition state of two Landau beams as given by
\begin{equation}
 \psi \ = \ \frac{1}{\sqrt{a^2 + b^2}} \,\left\{\, a \,\psi_{p_1, m_1}
 \ + \ b \,\psi_{p_2, m_2} \,\right\} .
\end{equation}
with two different values of the quantum numbers $p$ and $m$.
For simplicity, we assume here that the mixing coefficients $a$ and $b$ 
are both real numbers.
By using the relation $\Delta k_z (p, m) = - \,( 2 \,n + 1) \,/\,z_m$ with
$n \,=\, p + (|m| + m) / 2$ together with the relation
$z \,/\,z_m = \omega_L \,( z \,/\,v )$,
the probability density of the above superposition state is easily
calculated and given by
\begin{eqnarray}
 \rho &=& | \psi |^2 
 \ = \ \frac{1}{2 \,\pi (a^2 + b^2)} \,\left\{ \,
 a^2 \,N^2_{p_1, m_1} \,\left( \frac{2 \,r^2}{w^2_m} \right)^{|m_1|} \,
 e^{\,- \,\frac{2 \,r^2}{w^2_0}} \,
 \left[ L^{|m_1|}_{p_1} \left( \frac{2 \,r^2}{w^2_m} \right) \right]^2 \right. \nonumber \\
 &\,& \hspace{28mm}
 + \ b^2 \,N^2_{p_2, m_2} \,\left( \frac{2 \,r^2}{w^2_m} \right)^{|m_2|} \,
 e^{\,- \,\frac{2 \,r^2}{w^2_0}} \,
 \left[ L^{|m_2|}_{p_2} \left( \frac{2 \,r^2}{w^2_m} \right) \right]^2 \nonumber \\
 &\,& \hspace{8mm}
 + \ 2 \,a \,b \,N_{p_1, m_1} \,N_{p_2, m_2} \,
 \cos \left[ (m_1 - m_2) \,\left( \phi \ - \ 
 \frac{2 \,(n_1 - n_2)}{m_1 - m_2} \,\omega_L \,\,\frac{z}{v} \right) \,\right] \nonumber \\
 &\,& \hspace{26mm} \left. \ \times \ \ 
 \left( \frac{2 \,r^2}{w^2_m} \right)^{\frac{|m_1| + |m_2|}{2}} \,
 e^{\,- \,\frac{2 \,r^2}{w^2_0}} \,
 L^{|m_1|}_{p_1} \left( \frac{2 \,r^2}{w^2_m} \right) \, 
 L^{|m_2|}_{p_2} \left( \frac{2 \,r^2}{w^2_m} \right) \,\right\} , 
\end{eqnarray}
where
\begin{equation}
 N_{p_1, m_1} \ = \ \frac{2}{w_m} \,\sqrt{\frac{p_1 \,!}{(p_1 + |m_1|) \,!}}, \hspace{6mm}
 N_{p_2, m_2} \ = \ \frac{2}{w_m} \,\sqrt{\frac{p_2 \,!}{(p_2 + |m_2|) \,!}},
\end{equation}
and
\begin{equation}
 n_1 \ = \ p_1 \ + \ \frac{|m_1| + m_1}{2}, \hspace{6mm}
 n_2 \ = \ p_2 \ + \ \frac{|m_2| + m_2}{2} .
\end{equation}
Noticeable here is the appearance of the interference term, which
depends on the azimuthal angle $\phi$ as well as on the propagation 
distance $z$.
In consideration of the relation $z = v \,t$, the above expression
dictates that the interference term in the probability density rotates
around the $z$-axis with the angular velocity
\begin{equation}
 \overline{\omega} \ = \ \frac{2 \,(n_1 - n_2)}{m_1 - m_2} \,\,\omega_L .
 \label{Eq:average_angular_velocity}
\end{equation}
Here, we emphasize again the fact that, aside from these two
quantum numbers $m_1$ and $m_2$ specifying the incident beam,
$\overline{\omega}$ depends on $p_1, m_1$ and $p_2, m_2$ only through
the two quantum number $n_1$ and $n_2$ related to the Landau
quantum number \cite{BKMG2009}.

In the following analyses, we confine ourselves to simple cases in which the
numbers of node of the radial wave functions are both zero, i.e.
we limit to the cases in which $p_1 = p_2 = 0$ with the
arbitrary magnetic quantum numbers $m_1$ and $m_2$.
For the mixing coefficients $a$ and $b$, we shall use the
values given by 
$a = 1.0 \times \left( \sqrt{\,|m_1| \,!} \,/\,2^{\,|m_1| / 2} \right)$ and
$b = 2.0 \times \left( \sqrt{\,|m_2| \,!} \,/\,2^{\,|m_2| / 2} \right)$
throughout all the analyses below.
After setting $p_1 = p_2 = 0$, the above rotational
velocity $\overline{\omega}$ reduces to the form : 
\begin{equation}
 \overline{\omega} \ = \ \left(\, 1 \ + \ \frac{|m_1| - |m_2|}{m_1 - m_2} \,\right) \,\omega_L .
\end{equation}
Note that this expression is symmetric under the exchange of $m_1$ and $m_2$.
With use of the dimensionless radial coordinate $\xi = 2 \,r^2 / w^2_m$,
the probability density of the superposition state is then given by
\begin{eqnarray}
 \rho &=& \frac{1}{2 \,\pi} \,\,\frac{1}{a^2 + b^2} \,\,e^{\,- \,\xi} \,\,
 \Biggl\{ \,a^2 \,N^2_{0, m_1} \,\xi^{|m_1|} \ + \ 
 b^2 \,N^2_{0, m_2} \,\xi^{|m_2|} \Biggr. \nonumber \\
 &+& \ \Biggl. 2 \,a \,b \,N_{0, m_1} \,N_{0, m_2} \,\,
 \xi^{\frac{|m_1| + |m_2|}{2}} \,\cos \,\left[ (m_1 - m_2) \,
 \left(\, \phi \ - \ \frac{|m_1| + m_1 - |m_2| - m_2}{m_1 - m_2} \,\frac{z}{z_m} \right)
 \,\right] \, 
 \Biggr\} . \ \ \ \ \ 
\end{eqnarray}
We can also calculate the probability current density for the same
superposition state. For this state, the canonical part of the current contains 
both of radial and azimuthal components as
\begin{equation}
 \bm{j}^{\,can} \ = \ j^{\,can}_r \, \bm{e}_r \ + \ j^{\,can}_\phi \,\bm{e}_\phi,
\end{equation}
whereas the gauge part contains only the azimuthal component as
\begin{equation}
 \bm{j}^{\,gauge} \ = \ j^{\,gauge}_\phi \,\bm{e}_\phi .
\end{equation}
By using the dimensionless radial coordinate $\xi = 2 \,r^2 / w^2_m$,
they are expressed as
\begin{eqnarray}
 j^{\,can}_r &=& \frac{1}{2 \,\pi} \,\,\frac{a \,b}{a^2 + b^2} \,\,
 N_{0,m_1} \,N_{0,m_2} \,\,
 \frac{\sqrt{2}}{m_e \,w_m} \,\,
 \xi^{\,\frac{|m_1| + |m_2|}{2} - \frac{1}{2}} \,e^{\,- \,\xi} \nonumber \\
 &\,& \hspace{8mm} \times \ 
 \left( |m_1| - |m_2| \right) \,\sin \,\left[ (m_1 - m_2) \,
 \left(\, \phi \ - \ \frac{|m_1| + m_1 - |m_2| - m_2}{m_1 - m_2} \,\frac{z}{z_m} \,\right)
 \right], \hspace{21mm} \label{Eq:jcan_r}
\end{eqnarray}
\begin{eqnarray}
 j^{\,can}_\phi &=& \frac{1}{2 \,\pi} \,\,\frac{1}{a^2 + b^2} \,\,
 \frac{1}{m_e \,r} \,e^{\,- \,\xi} \,\,
 \Biggl\{ \,m_1 \,a^2 \,N^2_{0,m_1} \,\xi^{\,|m_1|} \ + \ 
 m_2 \,b^2 \, N^2_{0,m_2} \,\xi^{\,|m_2|} \Biggr. \nonumber \\
 &+& \!\!\! \Biggl. (m_1 + m_2 ) \,a \,b \,N_{0,m_1} \,N_{0,m_2} \,\,
 \xi^{\,\frac{|m_1| + |m_2|}{2}} \,\,\cos \,\left[ (m_1 - m_2) \,
 \left(\, \phi \ - \ \frac{|m_1| + m_1 - |m_2| - m_2}{m_1 - m_2} \,
 \frac{z}{z_m} \,\right) \,\right] \,\Biggr\} , \ \ \ \ \ \ \ \ \ 
\end{eqnarray}
and
\begin{eqnarray}
 j^{\,gauge}_\phi &=& \frac{1}{2 \,\pi} \,\,\frac{1}{a^2 + b^2} \,\,
 \frac{e \,B}{2 \,m_e} \,\,r \,\,e^{\,- \,\xi} \,\,
 \Biggl\{ \,a^2 \,N^2_{0,m_1} \,\xi^{\,|m_1|} \ + \ 
 b^2 \, N^2_{0,m_2} \,\xi^{\,|m_2|} \Biggr. , \nonumber \\
 &+& \Biggl. 2 \,a \,b \,N_{0,m_1} \,N_{0,m_2} \,\,
 \xi^{\,\frac{|m_1| + |m_2|}{2}} \,\,\cos \,\left[ \,(m_1 - m_2) \,
 \left(\, \phi \ - \ \frac{|m_1| + m_1 - |m_2| - m_2}{m_1 - m_2} \,
 \frac{z}{z_m} \, \right) \,\right] \,\Biggr\} . \hspace{10mm}
\end{eqnarray}

\begin{figure}[H]
\begin{center}
\includegraphics[width=11cm]{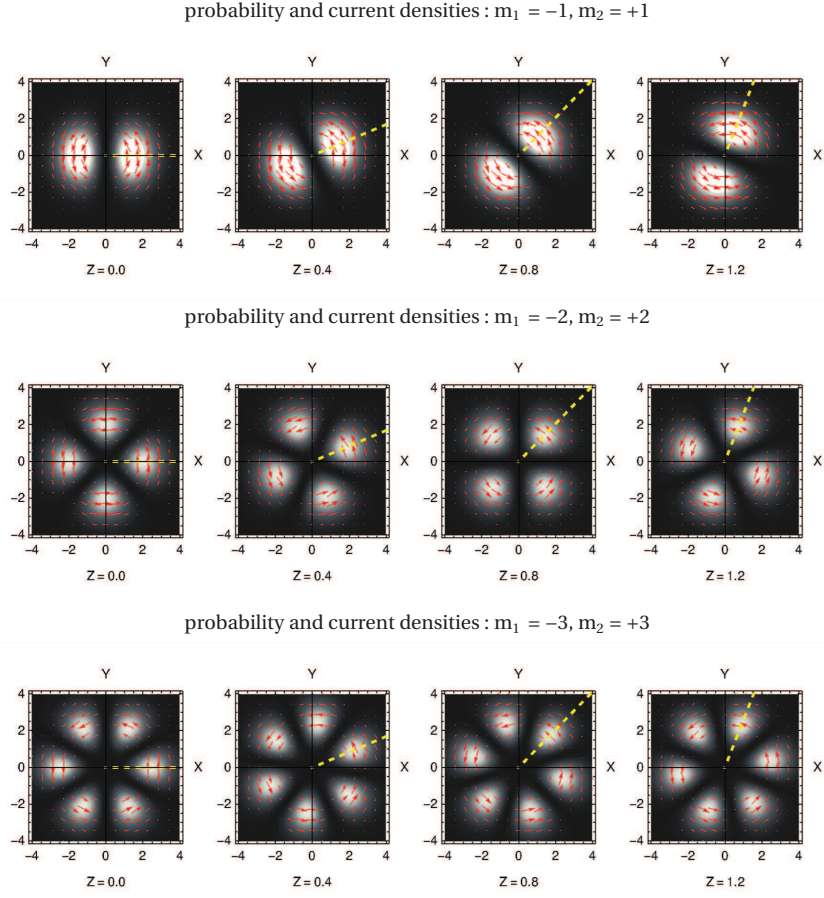}
\caption{The three (upper, middle, and  lower) panels represent the probability 
densities and the probability current densities in the $xy$-plane as functions of
the dimensionless propagation distance $Z = z / z_m$ for the OAM-balanced 
superposition state with $m_1 = - \,m_2$. 
Here and in the subsequent figures, the dimensionless coordinate $X$ and $Y$ 
are defined by $X \equiv x \,/\, l_B$ and $Y \equiv y \,/\, l_B$ with 
$l_B = w_m \,/\,2$. The upper, middle, and lower panels
respectively correspond to the states with
$(m_1, m_2) = (- \,1, + \,1)$, \,$(m_1, m_2) = (- \,2, + \,2)$, and
$(m_1, m_2) = (- \,3, + \,3)$.}
\label{Fig:Balanced_densities_BB} 
\end{center}
\end{figure}

Following \cite{BSVN2012}, we consider some typical cases in order to see
the effect of phase rotation in the superposed electron beam. 
The first is the OAM-balanced superposition
specified by $m_1 = - \,m_2 \equiv - \,m$ with $m$ being
arbitrary positive integer. Shown in Fig.\ref{Fig:Balanced_densities_BB} 
are the probability densities
as well as the probability current densities projected on the $xy$-plane 
as functions of the dimensionless propagation distance $Z = z / z_m$.
They are shown for four representative values of $Z$, i.e. 
$Z = 0, 0.4, 0.8$, and $1.2$.
The upper, middle and lower panels respectively correspond to
the superposition states with $(m_1, m_2)  = (- \,1, 1)$, \,
$(m_1, m_2) = (- \,2, 2)$, and $(m_1, m_2) = (- \,3, 3)$.

First, by looking at the leftmost three figures with $Z = 0$,
one clearly sees that the above three states respectively have
two, four and six brighter domains which corresponds to the
higher probability density regions. 
One can also see that, for all these three states, the probability 
current densities shown by
arrows (red in color) are circulating around the origin and
that they have no components along the radial direction.
This can be analytically convinced if one looks at the expression for
radial component of the current $j^{can}_r$ given in (\ref{Eq:jcan_r}).
The radial component is clearly seen to vanish in the OAM-balanced 
superposition with $|m_1| = |m_2|$.
As $Z$ increases, i.e. as the
beam propagates along the $z$-direction, one clearly sees the
rotation of the high density regions in the interference patterns.
One can readily check that the rotational velocity of these three superposition 
states with $(m_1, m_2)  = (- \,1, 1)$, \,$(m_1, m_2) = (- \,2, 2)$,
and $(m_1, m_2) = (- \,3, 3)$ are all the same and it is given by  
$\overline{\omega} = \left[ 1 +  (|m_1| - |m_2|) / (m_1 - m_2) \right]
\, \omega_L = \omega_L$, i.e. the Larmor frequency. This confirms
the observation in the paper \cite{BSVN2012}.

\begin{figure}[H]
\begin{center}
\includegraphics[width=12cm]{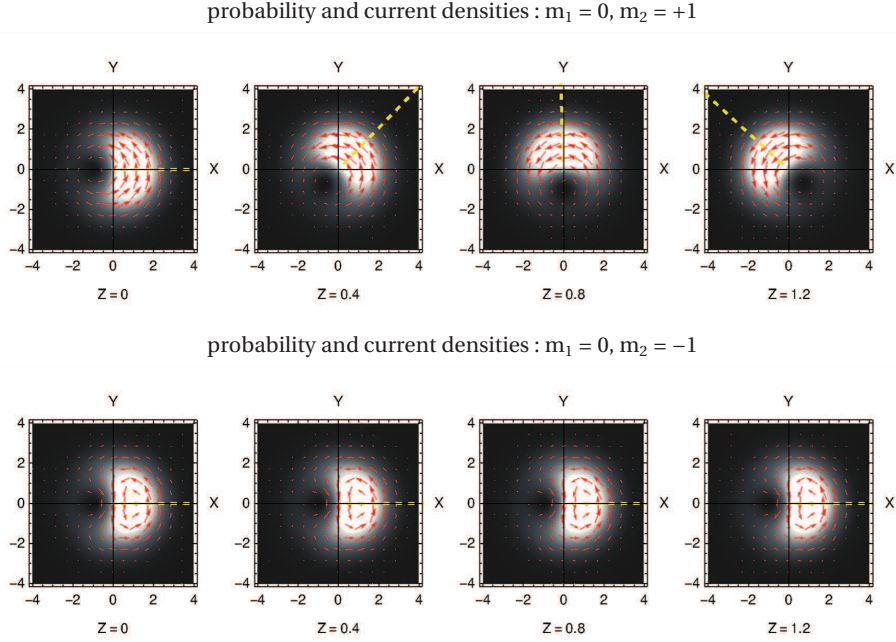}
\caption{The upper panel shows the probability densities and the current 
densities projected on the $xy$-plane for the superposition state with
$(m_1, m_2) = (0, + \,1)$ as functions of the dimensionless
propagation distance $Z = z / z_m$. 
Similar densities shown in the lower panel are those of the superposition
state with $(m_1, m_2) = (0, - \,1)$.}
\label{Fig:Unbalanced_Gray_0p1_0m1_BB} 
\end{center}
\end{figure}

Next, we consider the superposition in which one of the magnetic quantum 
number, say $m_1$ is zero, while another magnetic number $m_2$ is
arbitrary nonzero integer. Shown in Fig.\ref{Fig:Unbalanced_Gray_0p1_0m1_BB} 
are the probability densities
and the probability current densities projected on the $xy$-plane as
functions of the dimensionless propagation distance $Z$. 
The upper panel corresponds to the state with
$(m_1, m_2) = (0, + \,1)$, while the lower panel to the state with
$(m_1, m_2) = (0, - \,1)$. First, by looking at the leftmost two figures with
$Z = 0$, one confirms that the probability densities are exactly the
same for these two states with $(m_1, m_2) = (0, + \,1)$ and with 
$(m_1, m_2) = (0, - \,1)$.
However, the current densities for these two states show totally
different behaviors. The current for the state with $(m_1, m_2) = (0, + \, 1)$ is 
rotating counterclockwise and circulating the coordinate origin. 
On the other hand, the current
for the state with $(m_1, m_2) = (0, - \,1)$ is not circulating around the origin.
Rather, it is circulating around the center
of the high density region which is located on the positive $x$-axis.
This difference between the $m_2 > 0$ state and the $m_2 < 0$ state
also generates a remarkable dissimilarity in the rotational
behavior of the interference pattern. 
One clearly observes that, as $Z$ increases, the high density region
for the state with $(m_1, m_2) = (0, + \, 1)$ state rotates with a constant angular 
velocity. One can verify that this angular velocity is given by 
$\overline{\omega} = \left[\,1 +  (|m_1| - |m_2|) / (m_1 - m_2) \right] \,
\omega_L = 2 \,\omega_L = \omega_c$, i.e. the cyclotron frequency.  
In sharp contrast, the interference pattern for the superposition state with
$(m_1, m_2) = (0, - \,1)$ does not show any rotational behavior as $Z$ increases.
This is just consistent with the fact that the average rotational velocity of 
this state is given by 
$\overline{\omega} = \left[ 1 +  (|m_1| - |m_2|) / (m_1 - m_2) \right] \,\omega_L = 0$.

\begin{figure}[H]
\begin{center}
\includegraphics[width=12cm]{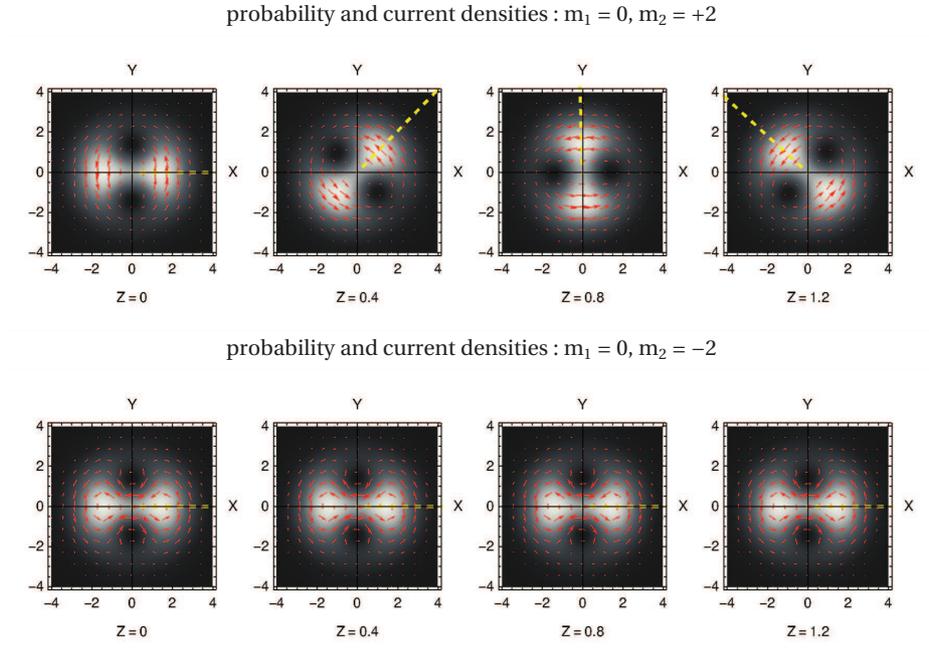}
\caption{The same as Fig.\ref{Fig:Balanced_densities_BB} 
except that the upper panel corresponds to the
superposition state with $(m_1, m_2) = (0, + \,2)$, while the lower panel
corresponds to the superposition state with $(m_1, m_2) = (0, - \,2)$.}
\label{Fig:Unbalanced_Gray_0p2_0m2_BB} 
\end{center}
\end{figure}

We also show in Fig.\ref{Fig:Unbalanced_Gray_0p2_0m2_BB} 
one more example of superposed beam
in which the first magnetic quantum number $m_1$ is zero but the
second magnetic quantum $m_2$ is nonzero. 
The upper panel of Fig.\ref{Fig:Unbalanced_Gray_0p2_0m2_BB} 
corresponds to the case with $(m_1, m_2) = (0, + \, 2)$, while the lower panel 
corresponds to the superposition 
state with $(m_1, m_2) = (0, - \,2)$. 
Again, by looking at the leftmost two figures with
$Z = 0$, one confirms that the probability densities of these two states
have two brighter regions and that probability densities of these two states
coincide perfectly with each other. 
However, the current densities of these two states
shown by arrows (red in color) are drastically different, and
this also generates totally different rotational behavior of the interference
patterns. One confirms that the interference pattern
for the state with $(m_1, m_2) = (0, + \, 2)$ rotates with the 
angular velocity $\omega_c$ when propagating along the $z$-direction.
On the other hand, the rotational velocity of the interference pattern
is zero for the superposition state with $(m_1, m_2) = (0, - \,2)$.

Aside from the number of peaks in the probability densities, all the
features of the states with $(m_1, m_2) = (0, \pm \,2)$ appear to be
nothing different from those of the states with $(m_1, m_2) = (0, \pm \,1)$.      
However, there is one profound difference between 
the states with $(m_1, m_2) = (0, \pm \,2)$ and the states with 
$(m_1, m_2) = (0, \pm \,1)$. In fact, making a careful inspection 
of Fig.\ref{Fig:Unbalanced_Gray_0p2_0m2_BB},
one realizes that the centroid of the former state with
$(m_1, m_2) = (0, \pm \,2)$ is always located at the coordinate center, but
this is not the case with the $(m_1, m_2) = (0, \pm \,1)$ states shown 
in Fig.\ref{Fig:Unbalanced_Gray_0p2_0m2_BB}.
At $Z = 0$, the centroid of the $(m_1, m_2) = (0, \pm \,1)$ states lies on 
the positive $x$-axis. When $Z$ is increased, 
the centroid of the $(m_1, m_2) = (0, + \,1)$ state rotates with the angular velocity 
$\omega_c$ around the $z$-axis, which means that the centroid of
this superposition state makes a spiral motion while propagating along the $z$-axis.
On the other hand, the centroid of the $(m_1, m_2) = (0, - \,1)$ state
makes no rotation in the $xy$-plane, because it remains on the same position
along $X$-axis. In other words, the centroid of this states makes 
a rectilinear motion along the $z$-axis. 

The helical motion of the centroid of the superposition state with $m_1 = 0$ 
and $m_2 = 1$ was already pointed out and called the ``spiralling of the 
Bohmian trajectory'' in the paper by 
Bliokh et al. \cite{BSVN2012},\cite{SSSLSBN2014}.
However, the description in their paper is a little misleading in that
it fails to emphasize the fact that $m_2 = \pm \,1$ mode is exceptional.
As a consequence, there might be a danger of giving a wrong impression 
that the superposition state with $m_1 = 0$ and arbitrary $m_2 \,(\,> 0)$ 
also makes a spiral motion.
Actually, as already pointed out above, the centroid of the 
$(m_1, m_2) = (0, \pm \,2)$ state is always located at the coordinate origin, 
so that it does not make a helical motion.
One can check that this is also the case for all the state with 
$m_1 = 0$ and $|m_2| = 2, 3, 4, \cdots$.
The reason why the $m_2 = \pm \,1$ modes are special can be understood
if one remembers the theory of nuclear deformation or that of the
electric dipole giant resonance in nuclei.
(See, for example, \cite{Book_Greiner1996}.)
For example, with use of the time-dependent shape parameter
$\alpha_{\lambda \mu} (t)$, the nuclear deformation is standardly 
parametrized in the following form 
$ R (\theta, \phi, t) \, = \, R_0 \,\left( \,1 \, + \, \sum_{\lambda = 0}^\infty \,
 \sum_{\mu = - \lambda}^\lambda \,\alpha^*_{\lambda \mu} (r) \,
 Y_{\lambda, \mu} (\theta, \phi) \,\right)$.
Here, $R (\theta, \phi, t)$ stands for the nuclear radius in the direction
$(\theta, \phi)$ at time $t$, while $R_0$ is a radius of spherical nucleus.
It is well-known that the $\lambda = 1$ mode is special among others.
In fact, the dipole deformations $\lambda = 1$ really do not
correspond to a deformation of the nucleus but rather to the {\it shift} 
of the center of mass of the nucleus. 
Similarly, in the present problem of electron beams,
one should notice that the wave function of the 
$m_2 = \pm \,1$ mode is proportional to $r \,e^{\,\pm \,\phi}
\propto r \,Y_{1, \pm \,1} (\theta = \frac{\pi}{2}, \phi)$.
This means that the mixture of the $m_2 = \pm \,1$ mode into
the $m_1 = 0$ mode works to shift the centroid of the superposition
beam toward the perpendicular direction to the beam axis.

Now we summarize in Table 1 the rotational velocity of the interference patterns
of the typical superposition states investigated so far. 
They are the OAM-balanced superposition with $m_1 = - \,m_2$
and the OAM-unbalanced superposition in which one of the magnetic quantum
number is zero. As already pointed out in \cite{BSVN2012}, three different rotational
velocities appear in these cases. 
They are the Larmor frequency $\omega_L$, the cyclotron
frequency $\omega_c$, and the zero frequency. Table 1 summarizes these
situations.


\begin{table}[htb]
\caption{The rotational velocities $\overline{\omega}$ of the 
OAM-balance superposition state with $(m_1 = + \,|m|, m_2 = - \,|m|)$ and also 
the OAM-unbalanced superposition states with 
$(m_1= 0, m_2= \mbox{any positive integer})$ and with
$(m_1= 0, m_2= \mbox{any negative integer})$.}  
\label{Table1}
\vspace{0mm}
\renewcommand{\arraystretch}{1.2}
\begin{center}
\begin{tabular}{cccc}
 \hline 
 \ \ \ $m_1$ \ \ \ \ \ & \ \ \ \ \ $+ \,|m|$ \ \ \ \ \ & \ \ \ 0 
 \ \ \ & \ \ \ 0 \ \ \ \\
 \hline
 \ \ \ $m_2$ \ \ \ \ \ & \ \ \ \ \ $- \,|m|$ \ \ \ \ \ &  \ \ \ positive integer
 \ \ \ & \ \ \ negative integer \ \ \ \\
 \hline
 \ \ \ $\overline{\omega}$ \ \ \ \ \ & \ \ \ \ \ $\omega_L$ \ \ \ & \ \ \ $\omega_c$ \ \ \ & 
 \ \ \ 0 \ \ \  \\
 \hline
\end{tabular}
\end{center}
\end{table}

\vspace{-2mm}
For the sake of completeness, let us consider some other types of OAM-unbalanced 
superpositions, which were not touched upon in \cite{BSVN2012}. 
Shown in Table 2 are the cases in which both
of $m_1$ and $m_2$ are positive integers or both of $m_1$ and $m_2$ are negative
integers. One sees that the superposition state with $m_1 > 0$ and $m_2 > 0$
rotates with the cyclotron frequency, while the superposition state with
$m_1 < 0$ and $m_2 < 0$ state shows no rotational behavior.
In any case, for the superposition state considered so far, only three types of
rotation frequency appear. They are the cyclotron frequency $\omega_c$, the 
Larmor frequency $\omega_L$, and zero frequencies.


\begin{table}[htb]
\caption{The rotational velocities $\overline{\omega}$ for the 
superposition state with
$(m_1=\mbox{any positive integer}, \,m_2=\mbox{any positive integer})$
and the state with 
$(m_1=\mbox{any negative integer}, \,m_2=\mbox{any negative integer})$.}
\label{Table2}
\vspace{0mm}
\renewcommand{\arraystretch}{1.2}
\begin{center}
\begin{tabular}{ccc}
 \hline 
 \ \ \ $m_1$ \ \ \ \ \ \ & \ \ \ \ \ positive integer \ \ \ \ \ & 
 \ \ \ \ \ negative integer \ \ \ \ \ \\
 \hline
 \ \ \ $m_2$ \ \ \ \ \ \ & \ \ \ \ \ positive integer \ \ \ \ \ &  
 \ \ \ \ \ negative integer  \ \ \ \ \ \\
 \hline
 \ \ \ $\overline{\omega}$ \ \ \ \ \ \ & \ \ \ \ \ $\omega_c$ \ \ \ \ \ & 
 \ \ \ \ \ 0 \ \ \ \ \ \\
 \hline
\end{tabular}
\end{center}
\end{table}


\begin{table}[htb]
\caption{The rotational velocities $\overline{\omega}$
for the superposed states with
$(m_1 = 1, m_2 = \mbox{any negative integer})$ and the
states with $(m_1 = - \,1, m_2 = \mbox{any positive integer})$.}
\label{Table3}
\vspace{0mm}
\renewcommand{\arraystretch}{1.2}
\begin{center}
\begin{tabular}{cccccccc}
 \hline 
 \ \ $m_1$ \ \ \ & \ \ + \,1 \ \ & \ \ + \,1 \ \ & \ \ + \,1 \ \ &
 \ \ + \,1 \ \ & \ \ + \,1 \ \ & \ \ $\cdots$ \ \ & \ \ + \,1 \ \ \\
 \hline
 \ \ $m_2$ \ \ \ & \ \ - \,1 \ \ & \ \ - \,2 \ \ &  
 \ \ - \,3 \ \ & \ \ - \,4 \ \ & \ \ - \,5 \ \ & 
 \ \ $\cdots$ \ \ & \ \ $- \,\infty$ \ \ \\
 \hline
 \ \ $\overline{\omega}$ \ \ \ & \ \ $\omega_L$ \ \ & 
 \ \ $\frac{2}{3} \,\omega_L$ \ \ & \ \ $\frac{2}{4} \,\omega_L$ \ \ & 
 \ \ $\frac{2}{5} \,\omega_L$ \ \ & \ $\frac{2}{6} \,\omega_L$ \ \ & 
 \ \ $\cdots$ \ \ & \ \ 0 \ \ \\
 \hline \\
 \hline 
 \ \ $m_1$ \ \ \ & \ \ - \,1 \ \ & \ \ - \,1 \ \ & \ \ - \,1 \ \ &
 \ \ - \,1 \ \ & \ \ - \,1 \ \ & \ \ $\cdots$ \ \ & \ \ - \,1 \ \ \\
 \hline
 \ \ $m_2$ \ \ \ & \ \ + \,1 \ \ & \ \ + \,2 \ \ &  
 \ \ + \,3 \ \ & \ \ + \,4 \ \ & \ \ + \,5 \ \ & 
 \ \ $\cdots$ \ \ & \ \ $+ \,\infty$ \ \ \\
 \hline
 \ \ $\overline{\omega}$ \ \ \ & \ \ $\omega_L$ \ \ & 
 \ \ $\frac{4}{3} \,\omega_L$ \ \ & \ \ $\frac{6}{4} \,\omega_L$ \ \ & 
 \ \ $\frac{8}{5} \,\omega_L$ \ \ & \ $\frac{10}{6} \,\omega_L$ \ \ & 
 \ \ $\cdots$ \ \ & \ \ $2 \,\omega_L$ \ \ \\
 \hline \\ 
\end{tabular}
\end{center}
\end{table}

Note however that, if one considers the superposition of LG beams
with $m_1 > 0$ and $m_2 < 0$ or with $m_1 < 0$ and $m_2 > 0$, 
one finds that the rotational dynamics
of the superposition state becomes far richer.
To see it, we investigate here two simple cases.
The first is the superposition state with $m_1 = 1$ and $m_2$
being any negative integer, while the second is the one
with $m_1 = - \,1$ and $m_2$ being any positive integer.  
As seen from Table 3, if one changes the value of $m_2$
from $- \,1$ to $- \,\infty$ with maintaning the value of $m_1$
to be $1$, the average rotational velocity $\overline{\omega}$ gradually
decreases and vanishes in the $m_2 \rightarrow - \,\infty$ limit.
On the other hand, Table 3 also shows that, if one changes the value of $m_2$ 
from $+ \,1$ to $+ \,\infty$ with maintaning the value of $m_1$ to be $- \,1$,
$\overline{\omega}$ gradually increases and approaches the cyclotron
frequency $\omega_c$ in the $m_2 \rightarrow + \,\infty$ limit.
Obviously, if one extends this analysis to more general cases with arbitrary 
value of $m_1$ and $m_2$
with $m _1 \,m_2 < 0$, wider variety of $\overline{\omega}$ in the
form of rational number times $\omega_L$ will appear.

\begin{figure}[H]
\begin{center}
\includegraphics[width=11cm]{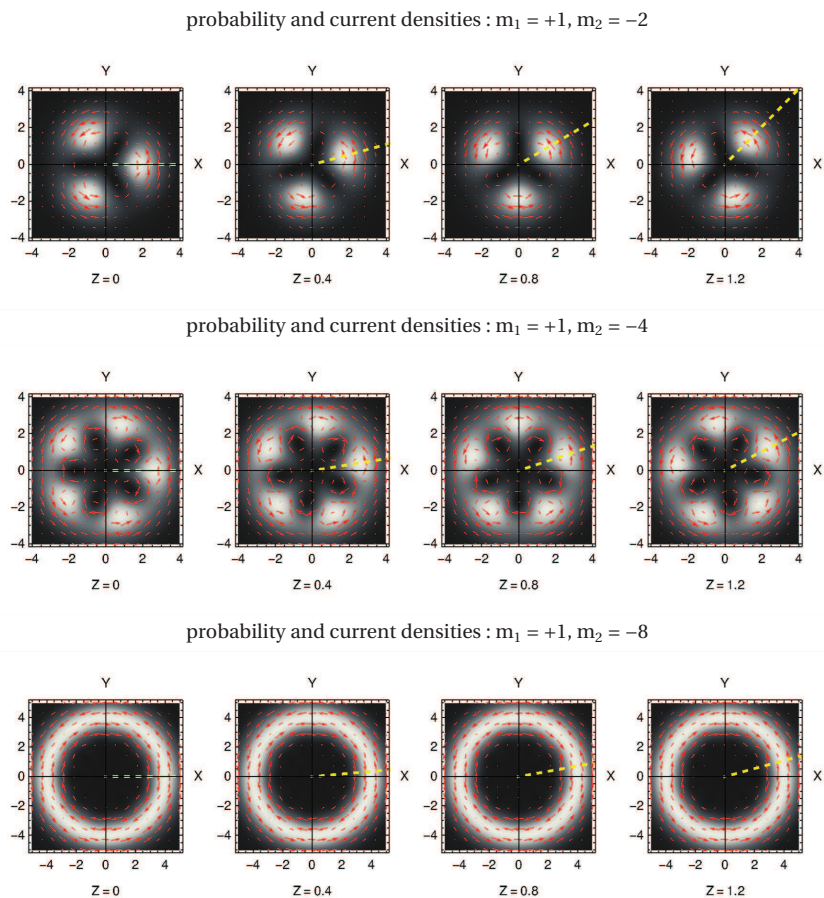}
\caption{The probability and current densities projected on the
$xy$-plane as functions of the dimensionless propagation distance
$Z = z / z_m$. The upper, middle, and lower panels respectively correspond
to the superposition states with $(m_1, m_2) = (+ \,1, - \,2)$, \,
$(m_1, m_2) = (+ \,1, - \,4)$, and $(m_1, m_2) = (+ \,1, - \,8)$.}
\label{Fig:Unbalanced_Gray_p1m248_BB} 
\end{center}
\end{figure}

To explicitly convince the above-explained structure of the rotational
dynamics of the superposition states, we show in 
Fig.\ref{Fig:Unbalanced_Gray_p1m248_BB} the probability
densities as well as the probability current densities projected on the 
$xy$-plane as functions of $Z = z / z_m$. 
The upper, middle, and the lower panels
respectively correspond to the superposition states with 
$(m_1, m_2) = (+ \,1, - \,2)$, $(m_1, m_2)  = (+ \,1, - \,4)$, and
$(m_1, m_2) = (+ \,1, - \,8)$. 
(In the last case with $(m_1, m_2) = (+ \,1, - \,8)$,  the density
distribution contains actually 9 blobs in the azimuthal direction,
but they are so close to each other that it looks like a ring.)

One clearly sees the rotations of
the interference patterns as functions of the dimensionless
propagation distance $Z$.
One can also convince that the rotational frequencies of these
three states are respectively given by  $\frac{2}{3} \,\omega_L$,
$\frac{2}{5} \,\omega_L$, and $\frac{2}{9} \,\omega_L$.
Clearly, $\overline{\omega}$ is approaching zero as $|m_2|$ increases.
 
\begin{figure}[H]
\begin{center}
\includegraphics[width=11cm]{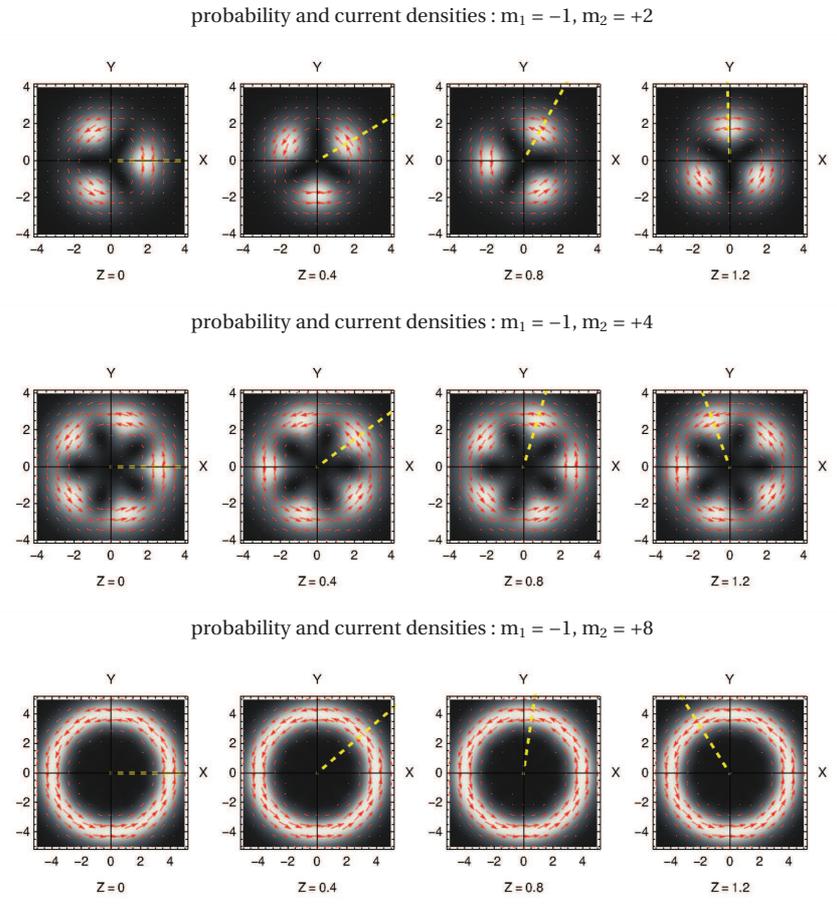}
\caption{The same as Fig.\ref{Fig:Unbalanced_Gray_p1m248_BB} 
except that the upper, middle, and lower 
panels respectively correspond to the superposition states with
$(m_1, m_2) = (- \,1, + \,2)$, \,$(m_1, m_2) = (- \,1, + \,4)$, and 
$(m_1, m_2) = (- \,1, + \,8)$.}
\label{Fig:Unbalanced_Gray_m1p248_BB} 
\end{center}
\end{figure}

Next, in Fig.\ref{Fig:Unbalanced_Gray_m1p248_BB}, we show the probability
densities as well as the probability current densities of the states with
$m_1 = - \,1$ and $m_2 = + \,2, + \,4, + \,8$, projected on the 
$xy$-plane as functions of $Z = z / z_m$. 
The upper, middle, and the lower panels
respectively correspond to the superposition states with 
$(m_1, m_2) = (- \,1, + \,2)$, and $(m_1, m_2) = (- \,1, + \,4)$,
$(m_1, m_2) = (- \,1, + \,8)$. 
(Just like the third figure in Fig.\ref{Fig:Unbalanced_Gray_p1m248_BB},
the density distribution in the third figure here actually contains 
9 blobs in the azimuthal direction.)
One again observes the rotations of
the interference patterns as $Z$ increases.
One can also convince that the rotational frequencies of these
three states are respectively given by  $\frac{4}{3} \,\omega_L$,
$\frac{8}{5} \,\omega_L$, and $\frac{16}{9} \,\omega_L$.
Clearly seen here is the tendency that 
$\overline{\omega}$ approaches $2 \,\omega_L = \omega_c$ 
as $|m_2|$ increases.

\vspace{-3mm}
In this way, we have seen that, if we consider the interference of the Landau
beams with two different magnetic quantum numbers $m_1$ and $m_2$,
there appear rich structures in the rotational dynamics of 
the superposition state. 
The emerging rotational frequency of the superposed probability
density is not limited to the cyclotron, Larmor, and zero frequencies.
Arbitrary rotational frequency in the form of rational number times $\omega_L$
can appear in general. These rotational velocities, which depend on 
the magnetic quantum number or the topological index of the beam,
are direct observables, since the rotation of the probability
density of the superposition state must be observed without any difficulty.
Our question here is the relation with observables in the Landau state,
especially the observability of the magnetic quantum number $m$
as the eigenvalue of the canonical OAM.
From the three-fold splitting of the rotational dynamics of the Landau states
as well as from the rotation of the interference pattern of the superposition
of the two beams, the authors of \cite{BSVN2012},\cite{SSSLSBN2014} 
gave an impression that the Larmor 
and the Gouy term of the total energy can separately be observed, which 
amounts to claiming that the eigenvalue of the canonical OAM is an observable.
On the other hand, however, it appears to contradict the fact that the 
canonical OAM in the Landau problem is a gauge-variant quantity.
The next section is devoted to the discussion of this very delicate and 
confusing problem.

\section{Discussions}
\label{Section5}


Throughout the investigations so far, motivated by the works of 
Bliokh et al.\cite{BSVN2012},\cite{SSSLSBN2014}, 
attention has been mostly paid to the similarity between the wave functions of the 
Landau states and those of the helical LG beams of the electron. 
Although the wave functions of these two systems are both described
by the Laguerre-Gauss type functions, one should not overlook a critical
difference between them. To understand the importance of this difference, 
let us first recall that, leaving 
aside the trivial free motion in the $z$-direction, the Landau Hamiltonian
is expressed as a sum of the Hamiltonian of the 2-dimensional harmonic 
oscillator and the Zeeman term. 
Naturally, the Hamiltonian of the 2-dimensional harmonic 
oscillator contains the restoring force potential, 
which is proportional to the square of the transverse distance 
from the coordinate origin. 
The cause of this restoring force term is of course the vector potential
in the symmetric gauge, which reproduces the uniform magnetic field along
the $z$-direction. To put it another way, what generates the cyclotron motion
of the Landau electron is the Lorentz force caused by the external magnetic
field. In sharp contrast, the helical LG beam of the electron is a solution
of the {\it free} Schr$\ddot{\text{o}}$dinger equation although in the paraxial 
approximation.
It may be a little perplexing that the wave function of the free electron
beam is expressed with the same functions as those of the Landau problem
with nonzero magnetic field.
The magic is the diffracting nature of the LG beam, which is embedded in
the $z$-dependence of the beam width. This $z$-dependence of the
beam width as well as that of the curvature radius of the wave front are the
key factors that the Laguerre-Gauss types of wave functions are
solutions of the free Schr$\ddot{\text{o}}$dinger equation in the
paraxial approximation. (This has been a long known fact in the field of twisted or helical 
photon beams in optics \cite{ABSW1992},\cite{APB1999},\cite{TT2011}.)

Let us now minutely compare the two OAMs, i.e. the canonical
and mechanical OAMs, with close attention to the above difference
between the non-diffractive Landau beams and the diffractive LG beams.      
We first recall that the magnetic quantum number $m$ (also called
the topological index) of the LG beam of the electron is widely believed to be a 
direct observable.
An immediate question is whether the observation of the canonical OAM
does not contradict the gauge principle or not, since we know that
the canonical OAM is believed to be a {\it gauge-variant} quantity.
A wayout of this seeming conflict may be found out as follows. 
First, note that, for free electrons, there is no substantial difference between the 
canonical OAM and the mechanical OAM because the gauge potential 
term is absent in a free space. 
Next, in the most general context, what appears in the dynamical equation of motion 
of a charged particle is the mechanical (or total) momentum or the mechanical (or total) 
OAM not the canonical momentum or canonical OAM. 
In a free space, however, the canonical OAM simply coincides with the total OAM, 
so that it is a quantity which appears in the equation of motion of the electron.
Then, it can in principle be extracted, for example, by measuring the torque exerted 
on the external atoms, etc. Undoubtedly, this would be
the reason why canonical OAM of the free LG beam can be a direct
observable without conflicting with the gauge principle.
Note, however, that the situation is a little different for the canonical OAM in the 
Landau problem.
Under the presence of the nonzero magnetic field, the canonical current
and the gauge current and also the canonical OAM and
the gauge part of the OAM are both nonzero, and besides they always appear as a 
single combination. As a consequence, the canonical part of the current or the OAM 
cannot be separately extracted, unless there are some external probes, 
which couples to the canonical current and the gauge current in a
different manner.

Now we recall that, based on the decomposition of the transverse part of the 
energy given by Eq.(\ref{Eq:E_Z_G}), Bliokh et al. argued that 
the Landau energy of an electron
in a magnetic field is given as the sum of the Zeeman and Gouy contributions as 
$E_\perp = E_Z + E_G$ \cite{BSVN2012}. 
They further claim that these two contributions 
are separately observable. This statement is a little misleading, however.
The statement would be certainly true for the diffractive LG beams, but it
is not necessarily true for the electron energy in the original Landau problem
or for the transverse energy in the nondiffractive Landau beam.
First, in the original Landau problem, the Zeeman part of the energy is proportional 
to the magnetic quantum number $m$ which is the eigen-value of the 
{\it gauge-variant} canonical OAM. Besides, as is clear from the discussion 
in sect.3, what-they-call the Gouy term of the energy (which is nothing but
the eigen-energy of the 2-dimensional harmonic oscillator) is also dependent 
on the choice of gauge. 
By this reason, it is usually believed that these two parts of energy are not
separately observable at least in the original Landau problem. 
The situation is basically the same also for the case of nondiffractive Landau
beam. As can be seen from (\ref{Eq:LZG_phase}), the Zeeman phase 
part proportional to $m$ and the Gouy phase part proportional to
$2 \,p + |m| + 1$ always appear in a single combination, so that they are be
separately measurable. 

The situation is quite different for the diffractive LG beam, however.
In fact, in the case of the diffractive LG beam, the Gouy term of the
transverse energy is connected with the Gouy phase of the diffractive
LG beam. 
Here, the Gouy phase of the diffractive LG beam takes the following form 
with $z_R$ being the Rayleigh diffraction length,
\begin{equation}
 \Phi_G \ = \ - \,(\,2 \,p \ + \ |m| \ + \ 1 \,) \,
 \arctan \left( \frac{z}{z_R} \right),
\end{equation}
which shows that the Gouy phase makes drastic jump before and behind the
beam waist where $z = 0$. This enables us to observe it directly and
separately with the Zeeman term.
This feature is of course a specific property of the diffractive LG 
beam of the electron not possessed by the nondiffractive Landau beam.
This after all means that the separate observation of the Zeeman term
and the what-they-call the Gouy term of the electron's energy
is a property of the diffracting LG beam but not a property of the 
Landau electron. 
For that reason, it would still be legitimate to say that the gauge-principle is not 
violated in the original Landau problem as well as in the physics of (free) helical
electron beams.

Finally, the difference between the non-diffractive Landau electron beam and 
the diffractive LG vortex electron beam was also discussed in a recent 
paper \cite{ZZS2021}. 
(See also the related paper \cite{ME2020}, which treats electron vortex beam 
in non-uniform magnetic field.)
In that paper, a general quantum-mechanical solution was obtained for the twisted 
electrons in a uniform magnetic field. This solution is remarkably different from
the nondiffractive Landau beam of Bliokh et al. and also from the free 
vortex beam of the electron. A prominent feature of 
this solution is that, as a reflection of its non-plane wave character
along the magnetic field direction, it correctly describes the $z$-dependence
of the beam width $w(z)$ for vortex electrons in a uniform magnetic field.
As a consequence, the angular velocity has no longer a simple relation 
with the Larmor frequency $\omega_L$ as given by (\ref{Eq:average_angular_velocity}), 
but it is generally $z$-coordinate dependent.
Another prominent feature of the above general vortex solution is that 
it coincides with Landau solution when condition $w_0 = w_m$ is fulfilled 
for the beam waist of the free vortex electron $w_0$ and the beam width
parameter $w_m$. If this condition is fulfilled, the beam radius becomes
a constant given by $w(z) = w_m = 2 \,l_B$. 
Under this circumstance, the electron angular velocity becomes $z$-independent 
and the conclusion obtained with Eq. (\ref{Eq:average_angular_velocity}) would be justified.
We also recall that, in \cite{SSSLSBN2014}, the electron beam was tuned such that 
the beam radius $w(z)$ comes close to magnetic radius $w_m$, i.e.
such that the relation $w(z) \simeq w_m$ holds. This might explain why the
experimental data in \cite{SSSLSBN2014} shows a good linear relation between
the angular velocity and the Larmor frequency.
As explained above, since the quantum-mechanical solutions of 
vortex electron beam and approximate solutions for the nondiffractive Landau 
beam are fairly different, a more careful analysis based on the exact 
solution would be highly desirable. It will be reported in a separated paper.

\section{Summary and conclusion}
\label{Section6}

Motivated by the observation of  the formal resemblance between the eigen-functions 
of the Landau problem and the wave functions of the diffractive LG beam,
Bliokh et al. \cite{BSVN2012} and also Schattschneider et al. \cite{SSSLSBN2014}
tried to disentangle the internal rotational dynamics of the Landau states, which
were not fully resolved in past studies of this old but popular problem.
Among others, they verified that the Landau modes with different 
magnetic (or azimuthal) quantum numbers $m$ belong to three classes,
which are characterized by rotations with zero, Larmor and cyclotron
frequencies, respectively. At first glance, in view of the fact that the magnetic 
quantum number $m$ of the Landau eigen-state is the eigen-value of the
{\it gauge-dependent} canonical OAM operator,
this $m$-dependent splitting of the Landau levels appears to expose a breakdown 
of the widely-accepted gauge principle. Motivated by this naive suspicion,
we have carried out a careful analysis of the 
$m$-dependent rotational dynamics of the Landau eigen-states
$|\,n, m \rangle$ in the symmetric gauge and confirmed that 
unexpectedly rich structure is hidden in the $m$-dependencies
of the Landau states.
First, we pointed out a novel symmetry of the electron's
probability densities of the two Landau states $|\,n-m, - \,m \rangle$
and $|\,n, m \rangle$. 
It was shown that these two states have exactly the same probability
densities, in spite that they have totally different eigen-energies.
The cause of this peculiar observation was verified to be traced
back to the difference between the probability current distributions of
these two states, which is generated by the interplay of the canonical and 
gauge-potential parts of the total (or mechanical) current, which critically
depends on the sign of the quantum number $m$. 
In any case, since this 3-fold splitting of the electron's rotational
velocity is a prediction based on the gauge-invariant total current (or
equivalently the mechanical current), our conclusion is that it does 
not contradict the gauge principle.

The conclusions up to this point were drawn chiefly based on the similarities 
between the Landau states and the wave functions of the single LG beams. 
If one considers superposition of the two Landau beams with different
magnetic quantum numbers $m_1$ and $m_2$, far richer structure turns out 
to appear in the rotational dynamics of such superposition states.
We begin with some simple cases already investigated in the paper by 
Bliokh et al. \cite{BSVN2012}. 
They are the OAM-balanced superposition with $m_1 = - \,m_2$,
the OAM-unbalanced superposition with $m_1 = 0$ and $m_2$ being some
positive or negative integers. In these cases, we confirmed their
observation that the rotational velocity of the superposed probability densities 
are limited to the three values, i.e. zero, Larmor and cyclotron frequencies.
However, if we consider more general superpositions with arbitrary
integers $m_1$ and $m_2$ with $m_1 \,m_2 < 0$, it turns out that wider variety 
of rotational velocity in the form of rational number times $\omega_L$
appear. A question is whether all these facts mean observations of the magnetic
quantum number $m$ in the Landau states. The answer is probably no. 
To convince it, we have emphasized a delicate but important difference 
between the non-diffractive Landau beam and the diffractive LG beam of
the electron. The former is a solution of a Schr$\ddot{\text{o}}$dinger equation 
under the presence of the nonzero magnetic field, while the latter is a solution of
a {\it free} Schr$\ddot{\text{o}}$dinger equation in a paraxial approximation.

Suppose now that, by following the experiments carried out by 
Schattschneider \cite{SSSLSBN2014},
the (free and diffractive) LG beam of the electron is incident into a nonzero
magnetic field region. 
As explained in the paper by Karlovets \cite{Karlovets2021}, which takes account 
of realistic experimental setup for preparing for magnetic fields, the free 
LG beam turns into nondiffractive Landau beam around the boundary of the 
free space and the nonzero magnetic field region. This transition happens
in a finite transition period, i.e. in the cyclotron period.
Once entering the nonzero magnetic field region, the phase
rotation of the nondiffractive Landau beam is determined by
$m + (\,2 \,p + |m| + 1) = 2 \,n + 1$, so that the Larmor and Gouy 
part of the phase rotation cannot be separated.
However, after going out of the nonzero magnetic field region,
the nondiffractive Landau beam again turns into the free LG beam.
After propagating a long distance, the $z$-dependence of the width of
this free LG beam becomes appreciable and the beam eventually reaches 
the focal point where the drastic change of the Gouy phase happens.
This would make it possible to separate the 
Zeeman and Gouy part of the phase rotation.
We therefore think that the separation of the corresponding Zeeman and Gouy
parts of the electron's transverse energy should be thought of as the property
of the diffractive LG beam, and it should not be interpreted as $m$-dependent
splitting of the Landau state.
If we think this way, we can say that any of the 
observations made in the papers \cite{BSVN2012},\cite{SSSLSBN2014}
would not contradict the widely-accepted gauge principle.

\noindent
\section*{Acknowledgement}


\vspace{2mm}
\noindent
M.~W. thanks the Institute of Modern Physics of the Chinese
Academy of Sciences in Lanzhou for hospitality.
Y.~K. L-P.~.Z and P.-M.~Z. are supported by the National Natural
Science Foundation of China (Grant No.11975320 and No.11805242).
This work is partly supported by the Chinese Academy of Sciences
President's International Fellowship Initiative 
(No. 2018VMA0030 and No. 2018PM0028).



\vspace{8mm}
\noindent

\end{document}